\documentclass[aip,pof,reprint,amsmath,amssymb]{revtex4-2}

\usepackage[utf8]{inputenc}
\usepackage[T1]{fontenc}
\usepackage{lmodern}
\usepackage{bm}
\usepackage{graphicx}
\usepackage{booktabs}
\usepackage{placeins}
\usepackage[colorlinks=true,linkcolor=blue,citecolor=blue,urlcolor=blue]{hyperref}

\newcommand{\bu}{\bm{u}}
\newcommand{\bE}{\bm{E}}
\newcommand{\bv}{\bm{v}}
\newcommand{\ua}{\bm{u}^{\dagger}}   %
\newcommand{\pa}{p^{\dagger}}        %
\newcommand{\qa}{q^{\dagger}}        %
\newcommand{\phia}{\phi^{\dagger}}   %
\newcommand{\bz}{\bm{z}}
\newcommand{\bn}{\bm{n}}
\newcommand{\dd}{\,\mathrm{d}}

\begin{document}
\makeatletter\immediate\write\@auxout{\string\citation{aipcontrol}}\makeatother%

\title{Nonlinear non-modal stability and minimal perturbations in
electroconvection driven by unipolar injection}

\author{Pedro A. V\'azquez}
\email{pvazquez@us.es}
\affiliation{Department of Applied Physics III, ETSI, Universidad de Sevilla,
Camino de los Descubrimientos s/n, 41092 Sevilla, Spain}

\author{A. T. P\'erez}
\affiliation{Department of Electronics and Electromagnetism, Faculty of Physics,
Universidad de Sevilla, Avda. Reina Mercedes s/n, 41012 Sevilla, Spain}

\begin{abstract}
Electroconvection in a dielectric liquid subjected to unipolar charge
injection is subcritical. Convection can be sustained well below the
threshold of linear stability theory, and experiments report onset far below
it. We use nonlinear non-modal optimization to determine the smallest charge
perturbations that trigger convection below the linear threshold. The method,
based on a direct--adjoint formulation of the full electrohydrodynamic
equations, recovers the classical linear threshold in the limit of small
amplitudes. At finite amplitude, the most dangerous perturbations are
localized charge deficits rather than excesses, because a deficit relaxes
more slowly under Coulomb repulsion. The minimal perturbation needed for
transition grows from about 0.5\% of the injected charge near the linear
threshold to 3\% at two thirds of it, where the convective branch ends, in
agreement with direct numerical simulations. Near the critical amplitude,
trajectories approach an unstable steady state, the edge state, that
separates the basins of attraction of rest and convection and originates
from the critical linear mode. The damped oscillatory modes involved in the
growth of the optimal perturbations are transit-time modes of the ion drift.
The convective mode is the first of them, made stationary by the coupling
with the flow, which explains why the instability sets in as a steady
motion. Small charge fluctuations can therefore lower
the onset of electroconvection to about two thirds of the linear threshold, 
helping to explain the discrepancy between linear theory and experiments.
\end{abstract}

\maketitle

\section{Introduction}

Electrohydrodynamic (EHD) flows arise from the interaction of electric fields
with electric charges in a dielectric fluid \citep{Castellanos1998}.
Electroconvection driven by unipolar charge injection is an
electrohydrodynamic instability in which Coulomb forces generated by space
charge can destabilize an otherwise motionless dielectric liquid. One of the
most studied configurations consists of a horizontal layer of dielectric
liquid confined between two parallel electrodes, with a strong electric field
applied across the layer and unipolar charge injection at one of the
electrodes. The electric field acts on the injected space charge, generating
a body force that can drive fluid motion and lead to the formation of
vortical structures known as electroconvective rolls.

While this flow bears a strong geometric and physical resemblance to thermal
Rayleigh--B\'enard convection, its charge transport mechanisms lead to
fundamentally distinct bifurcation dynamics. The stability of the flow is
governed by a dimensionless parameter, the electric Rayleigh number
$T=\varepsilon V_0/(\mu K)$, which represents the ratio of electric to viscous
forces, where $\varepsilon$ is the permittivity of the fluid, $V_0$ is the
applied voltage, $\mu$ is the dynamic viscosity and $K$ is the ionic mobility.
The critical value of this parameter, denoted $T_c$, marks the onset of
electroconvective motion.

Early experiments had already reported electrically driven vortical motion in
dielectric liquids, with observations dating back at least to
\citet{Avsec1937}. Measurements by \citet{Watson1970} confirmed the onset of
electroconvective motion and current enhancement in insulating liquids under
space-charge-limited conditions. Subsequent experiments by
\citet{Lacroix1975} described convective cell velocity fields and verified
the subcritical hysteresis loop, with convection setting in near
$T\approx100$, well below the linear prediction. Current fluctuation measurements by \citet{Malraison1982}
further demonstrated that in moderate-to-large aspect ratio cells the flow
undergoes a direct transition from rest into time-dependent chaotic motion,
shifting from a viscous-dominated regime at moderate voltages to an
inertially dominated regime at high voltages. Later transient experiments by
\citet{Pontiga1995} broadened this empirical foundation to non-autonomous,
field-dependent charge injection in salt-doped nonpolar liquids by tracking
step-voltage responses and apparent ionic mobility.

Linear stability analysis of unipolar charge injection between parallel plate
electrodes was established by \citet{SchneiderWatson1970} and
\citet{AttenMoreau1972}, who formulated the governing equations. Under
space-charge-limited (SCL) autonomous injection, classical modal stability
theory predicts that the hydrostatic rest state loses stability to infinitesimal
disturbances at a critical threshold $T_c\approx161$ (or $160.75$)
\citep{SchneiderWatson1970}, whereas in the weak injection regime the marginal
criterion scales as $T_cC^2\approx220.75$. Here, $C=q_iH^2/(\varepsilon V_0)$ is the
injection level parameter, where $q_i$ is the charge density at the injector and
$H$ is the electrode separation. \citet{Pontiga1995} extended
this linear framework to non-autonomous, field-dependent charge injection
(governed by the Schottky effect), demonstrating that while field-dependent
boundary conditions noticeably alter marginal stability at moderate injection
levels, the threshold in the strong injection limit remains $T_c\approx161$.
Although classical models neglected charge diffusion because of its small
magnitude relative to ionic drift, subsequent analyses by
\citet{PerezCastellanos1989} and \citet{Zhang2015} showed that incorporating
realistic charge diffusion smooths charge density gradients and lowers the
theoretical linear threshold to $T_c\approx140$.

Non-modal stability theory provides an alternative to the classical
eigenvalue approach by focusing on the finite-time amplification of
perturbations rather than on their asymptotic exponential growth. In linear
systems this approach is particularly relevant when the linearized evolution
operator is non-normal, since a superposition of decaying eigenmodes may
undergo substantial transient amplification
\citep{Trefethen1993,Luchini2014}. The corresponding optimal perturbation is
defined as the initial disturbance, subject to a prescribed norm, that
maximizes a chosen measure of the perturbation at a specified target time. In
fluid mechanics, linear optimal perturbations and transient growth have
become standard tools for identifying the disturbance structures that are
most efficiently amplified by the underlying flow dynamics.

The same optimization concept can be extended to finite-amplitude
disturbances by retaining the nonlinear terms in the governing equations. In
this case the optimal perturbation is obtained by maximizing a finite-time
objective functional subject to the full nonlinear evolution equations and to
a prescribed initial disturbance amplitude. The resulting constrained
optimization problem is commonly solved by a direct--adjoint looping
procedure. An initial perturbation is first evolved forward in time, the
associated adjoint equations are then integrated backward from the target
time, and the initial condition is iteratively updated in the direction of
the gradient of the objective functional. This nonlinear extension of
transient-growth analysis was developed in the context of transition to
turbulence by \citet{PringleKerswell2010} and \citet{Cherubini2011}, and was
subsequently used to identify finite-amplitude optimal perturbations and
minimal seeds for transition \citep{Pringle2012,Kerswell2018}. More recently,
the framework has been extended to other forms of finite-amplitude
optimization, including sparse nonlinear optimal perturbations
\citep{Heide2025}.

Although nonlinear non-modal optimization has been applied extensively to
hydrodynamic transition problems, its application to electrohydrodynamic
instability remains comparatively unexplored. A first indication of the
possible role of non-modal effects in unipolar-injection electroconvection was
provided by \citet{Zhang2015}, who performed a linear transient-growth
analysis of the problem. For the configuration without cross-flow they found
a finite but relatively small transient amplification, insufficient to
account for the discrepancy between the linear stability threshold and the
lower finite-amplitude onset observed experimentally. This suggests that the
essential mechanism responsible for subcritical electroconvection cannot be
captured within an infinitesimal, linearized description. A different
explanation was explored by \citet{Feng2021}, who attributed the discrepancy to
the inhomogeneity of the ion-exchange membranes used as injectors in the
experiments. Modeling it with stochastic boundary conditions for the charge
density and the electric potential, they found that, owing to the subcritical
nature of the flow, the threshold for the onset of convection decreases as the
amplitude of the fluctuations increases and approaches the experimental values.
The present work quantifies, with an optimization approach, the smallest
charge perturbation that can trigger convection below the linear threshold.

\begin{figure*}
  \centering
  \includegraphics[width=\linewidth]{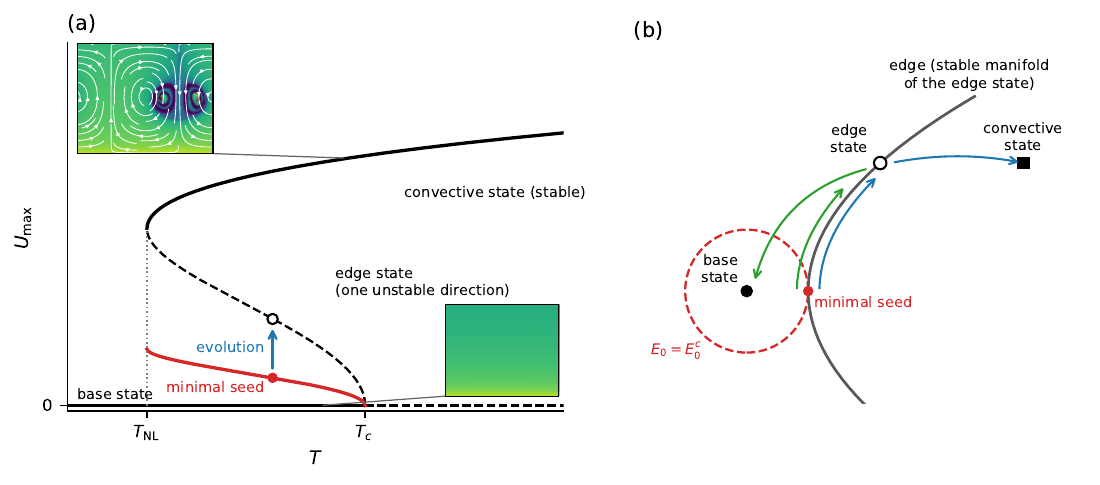}
  \caption{ Sketch of the subcritical transition. (a) Bifurcation diagram in
    terms of the maximum velocity $U_{\max}$, with stable branches (solid) and
    edge states (dashed). The red curve is only illustrative. It is not a state,
    and indicates that the minimal seed lies on the stable manifold of the edge
    state. Insets: total charge $q_b+q'$ (logarithmic color scale, as in
    Fig.~\ref{fig:edge_states}) in a convective state at $T=112.23$, close to
    $T_{\rm NL}$, with streamlines, and in the hydrostatic state, which has no rolls. (b) Phase space
    at fixed $T_{\rm NL}<T<T_c$. The edge (grey) separates the basins of the
    base and convective states, and the sphere $E_0=E_0^c$ (dashed red) touches it
    at the minimal seed. Blue and green: perturbations just above and just below
    $E_0^c$. The curves are qualitative. }
  \label{fig:sketch}
\end{figure*}

Figure~\ref{fig:sketch} summarizes the picture that guides this work. The
unipolar-injection problem is subcritical. Between a nonlinear threshold $T_{\rm
NL}$ and $T_c$ the stable hydrostatic state coexists with a stable convective
state [Fig.~\ref{fig:sketch}(a)]. The two are separated by an unstable branch of
edge states (dashed line), which bifurcates from the hydrostatic state at $T_c$
and meets the stable convective branch at a saddle-node bifurcation at $T_{\rm
NL}$. In phase space [Fig.~\ref{fig:sketch}(b)], a state on this unstable branch
\citep{Skufca2006,Schneider2007} has a single unstable direction, which leads
either to rest or to convection. Its stable manifold, the edge, separates the
basins of attraction of the two stable states. The minimal seed is the point of
the edge closest to the hydrostatic state, i.e., the perturbation of smallest
amplitude that can trigger transition \citep{Pringle2012,Kerswell2018}, and the
sphere $E_0=E_0^c$ (dashed circle) touches the edge there. Its amplitude
$E_0^c(T)$ measures the robustness of the hydrostatic state. In
Fig.~\ref{fig:sketch}(a) the branches are represented by the maximum velocity
$U_{\max}$, which identifies them unambiguously. The hydrostatic state has
$U_{\max}=0$. The insets show the charge distribution in the hydrostatic
state, which is stratified and has no rolls, and in a convective state, with
two counter-rotating rolls. The red curve is only illustrative. It is not a
state, since the minimal seed is a charge perturbation with zero initial
velocity. It indicates that, for every $T$ between $T_{\rm NL}$ and $T_c$, the
minimal seed lies on the stable manifold of the edge state. Smaller perturbations decay
directly. Perturbations just above (blue) and just below (green) $E_0^c$ follow
the edge to the neighborhood of the edge state and leave it along its unstable
manifold, towards convection or back to rest. The minimal seed itself would
converge to the edge state. The curves are qualitative, and
Fig.~\ref{fig:branches} shows the computed diagram.

The nonlinear optimization problem is discretized with Firedrake
\citep{Rathgeber2016}, a finite-element library. The corresponding discrete adjoint
equations and gradients are generated automatically with the Python library
Pyadjoint \citep{Mitusch2019}, which succeeds the dolfin-adjoint library
\citep{Farrell2013}. It exploits the high-level variational representation of
the forward problem to construct discrete tangent-linear and adjoint models
directly from the finite-element formulation, avoiding the need to derive and
implement the discrete adjoint system by hand. 

The paper is organized as follows. Section~\ref{sec:formulation} presents the
governing equations of unipolar-injection electroconvection in a periodic
layer, the hydrostatic base state and the classical linear modal problem,
whose discrete threshold serves as the reference for the rest of the work.
Section~\ref{sec:optimization} formulates the nonlinear non-modal
optimization problem and describes its discretization and the
direct--adjoint optimization procedure. Section~\ref{sec:validation} assesses
the spatial and temporal convergence of the method, determines the amplitude
below which the optimization behaves linearly, and shows that the non-modal
framework recovers the linear instability threshold. Section~\ref{sec:nonlinear}
contains the main results. These are the finite-amplitude optimal perturbations below
the linear threshold and their structure, the nonlinear mechanism that
favors charge-deficit perturbations, the minimal seed for transition to the
convective state, the edge state and the end of the convective branch
$T_{\rm NL}$, compared with direct numerical simulations, and a
decomposition of the optimal trajectories on direct and adjoint temporal
eigenmodes. The implications for the onset of
electroconvection are discussed in Sec.~\ref{sec:onset}.
Section~\ref{sec:transit} identifies the damped oscillatory modes revealed by
this decomposition as transit-time modes of the ion drift, shows that the
convective mode is the first of them made stationary by the flow, and
verifies that none of them becomes unstable. Conclusions are
drawn in Sec.~\ref{sec:conclusions}. The derivation of the adjoint equations
is given in Appendix~\ref{app:adjoint}, the monolithic weak formulation
of the direct problem in Appendix~\ref{app:weak}, and the exact solution for
the horizontally uniform transit-time modes in Appendix~\ref{app:transit}.

\section{Governing equations and linear stability reference}
\label{sec:formulation}

\subsection{Dimensional equations}

We consider a horizontal layer of dielectric liquid of thickness $H$ confined
between two parallel electrodes, with a strong electric field applied across
the layer and unipolar charge injected at the lower electrode. The governing
equations for the fluid motion, charge transport and electric field are the
incompressible Navier--Stokes equations coupled with charge conservation and
Poisson's equation:
\begin{align}
  \rho\left(\frac{\partial\bu}{\partial t}+\bu\cdot\nabla\bu\right)
    &= -\nabla p+\mu\nabla^2\bu+q\bE, \label{eq:ns_dim}\\
  \nabla\cdot\bu &= 0, \label{eq:div_dim}\\
  \frac{\partial q}{\partial t}+\nabla\cdot\left[q\,(\bu+K\bE)\right] &= 0,
    \label{eq:charge_dim}\\
  \nabla^2\phi = -q/\varepsilon, \qquad \bE &= -\nabla\phi .
    \label{eq:poisson_dim}
\end{align}
Here $\bu$ is the fluid velocity, $p$ the pressure, $\phi$ the electric
potential, $q$ the electric charge density, $\bE$ the electric field,
$\varepsilon$ the electric permittivity of the fluid, $\rho$ the fluid mass density,
$\mu$ the dynamic viscosity and $K$ the ionic mobility. The electric force is
due only to the Coulomb force $q\bE$, as the liquid is assumed isothermal. The
electrostriction force, being the gradient of a scalar function, can always be
absorbed into the pressure. In this study, we analyze how the non-modal and
nonlinear aspects of the system affect the predictions of the classical linear
stability analysis, which does not include charge diffusion. Therefore, we
neglect charge diffusion in Eq.~\eqref{eq:charge_dim}.

\begin{figure*}
  \centering
  \begin{minipage}[t]{0.49\textwidth}
    \raggedright\textsf{(a)}\\[2pt]
    \includegraphics[width=\linewidth]{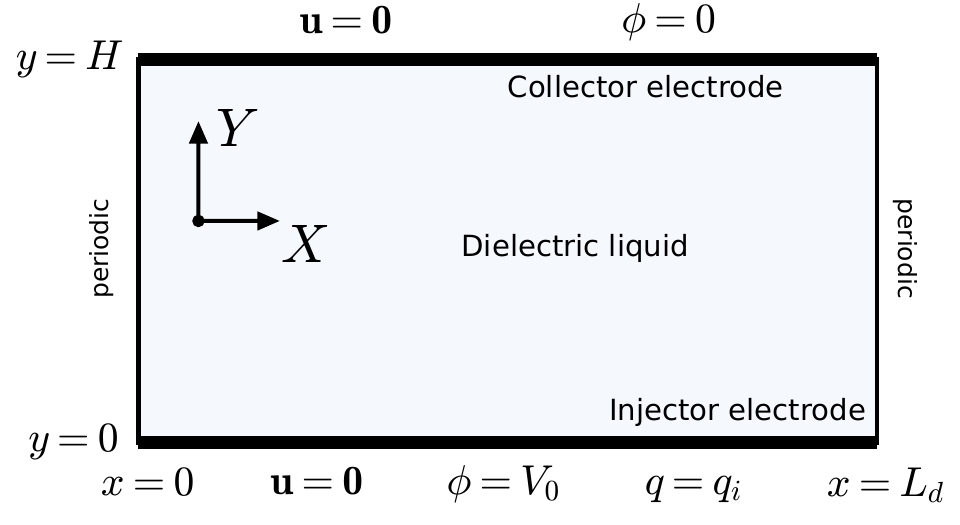}
  \end{minipage}\hfill
  \begin{minipage}[t]{0.49\textwidth}
    \raggedright\textsf{(b)}\\[2pt]
    \includegraphics[width=\linewidth]{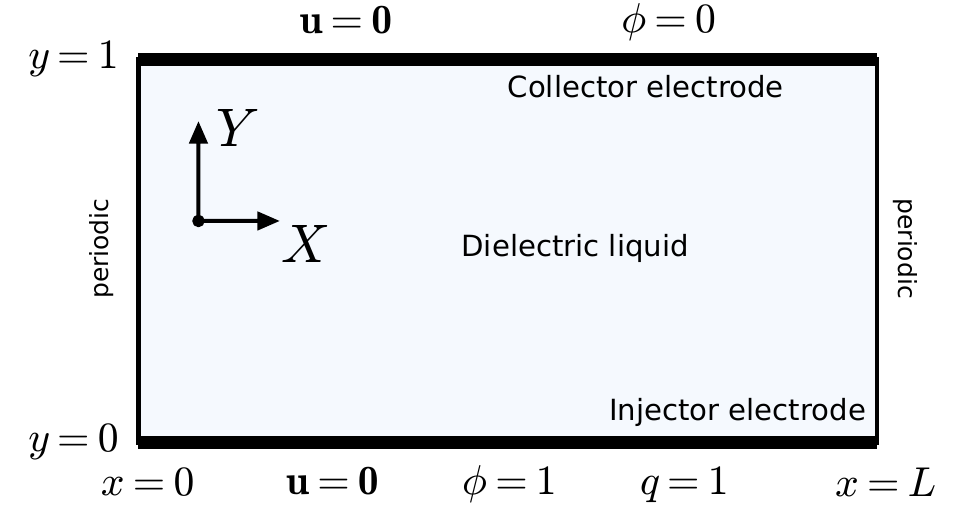}
  \end{minipage}
  \caption{Domain geometry and boundary conditions for the electroconvection
    problem. (a) Dimensional. (b) Nondimensional.}
  \label{fig:domain_dim}
\end{figure*}

The boundary conditions [Fig.~\ref{fig:domain_dim}(a)] are
\begin{equation}
  \begin{aligned}
    y=0 &: \quad \bu=\bm{0},\quad \phi=V_0,\quad q=q_i,\\
    y=H &: \quad \bu=\bm{0},\quad \phi=0,\\
    &\phantom{:}\quad \bu,\ p,\ q,\ \phi \ \text{periodic in } x
      \text{ with period } L_d.
  \end{aligned}
  \label{eq:bc_dim}
\end{equation}
Here $q_i$ is the charge density at the injector electrode, $V_0$ is the
applied voltage and $L_d$ is the lateral period of the domain. At $y=0$ we
impose no slip, a fixed potential and a prescribed injected charge density.
At $y=H$ we impose no slip and a grounded potential. In the horizontal
direction the layer is periodic, which models a laterally unbounded layer
and allows perturbations of any horizontal phase. The charge conservation
equation, Eq.~\eqref{eq:charge_dim}, is hyperbolic, so a charge boundary condition
is needed only where the charge transport velocity $\bu+K\bE$ enters the
domain, i.e., at the injector. At the collector the charge leaves the domain
and no condition is imposed.

Equations~\eqref{eq:ns_dim}--\eqref{eq:poisson_dim} have a steady base state
corresponding to the hydrostatic conduction state,
\begin{equation}
  \bu_b=\bm{0},\qquad q_b=q_b(y),\qquad \phi_b=\phi_b(y),
  \label{eq:base_def}
\end{equation}
where $q_b(y)$ and $\phi_b(y)$ are the base charge density and electric
potential profiles. They are determined by the steady versions of
Eqs.~\eqref{eq:charge_dim} and \eqref{eq:poisson_dim}, which reduce to
\begin{equation}
  \frac{\dd E_b}{\dd y}=\frac{q_b}{\varepsilon},\qquad
  \frac{\dd}{\dd y}\left(q_bE_b\right)=0 .
  \label{eq:base_ode}
\end{equation}
Their solution is
\begin{equation}
  \begin{split}
  &E_b(y)=E_i\sqrt{1+\frac{2q_i}{\varepsilon E_i}\,y},\\
  &q_b(y)=q_i\left(1+\frac{2q_i}{\varepsilon E_i}\,y\right)^{-1/2},
  \end{split}
  \label{eq:base_dim}
\end{equation}
where $E_i$ is the electric field at the injector electrode. It can be computed
imposing that the potential drop across the layer equals the applied voltage:
\begin{equation}
\begin{split}
  &V_0=\int_0^H E_b(y)\dd y
  \quad\Longrightarrow\quad \\
  &E_i^2\left[\left(1+\frac{2q_iH}{\varepsilon E_i}\right)^{3/2}-1\right]
  =\frac{3q_iV_0}{\varepsilon}.
  \label{eq:Es_dim}
\end{split}
\end{equation}
This equation is solved numerically for $E_i$, after which the base-state
profiles $E_b(y)$ and $q_b(y)$ follow from Eq.~\eqref{eq:base_dim}.

\subsection{Nondimensional equations and parameters}

We choose the characteristic scales
\begin{equation}
  \begin{aligned}
  &x,y\sim H,\quad    & t\sim\frac{H^2}{KV_0},\quad &u\sim\frac{KV_0}{H},\quad p\sim\frac{\rho K^2V_0^2}{H^2},\\
  &\phi\sim V_0,\quad & E\sim\frac{V_0}{H},\quad    &q\sim q_i .
  \label{eq:scales}
  \end{aligned}
\end{equation}
The nondimensional version of Eqs.~\eqref{eq:ns_dim}--\eqref{eq:poisson_dim} is
\begin{align}
  \frac{\partial\bu}{\partial t}+\bu\cdot\nabla\bu
    &= -\nabla p+\frac{M^2}{T}\nabla^2\bu+CM^2q\bE, \label{eq:ns}\\
  \nabla\cdot\bu &= 0, \label{eq:div}\\
  \frac{\partial q}{\partial t}+\nabla\cdot\left[q\,(\bu+\bE)\right] &= 0,
    \label{eq:charge}\\
  \nabla^2\phi = -Cq,\qquad \bE &= -\nabla\phi . \label{eq:poisson}
\end{align}
From now on all variables are nondimensional unless otherwise stated. The
nondimensional boundary conditions [Fig.~\ref{fig:domain_dim}(b)] are
\begin{equation}
  \begin{aligned}
    y=0 &: \quad \bu=\bm{0},\quad \phi=1,\quad q=1,\\
    y=1 &: \quad \bu=\bm{0},\quad \phi=0,\\
    &\phantom{:}\quad \bu,\ p,\ q,\ \phi \ \text{periodic in } x
      \text{ with period } L .
  \end{aligned}
  \label{eq:bc}
\end{equation}

The nondimensional parameters are
\begin{equation}
  T=\frac{\varepsilon V_0}{\mu K},\qquad
  C=\frac{q_iH^2}{\varepsilon V_0},\qquad
  M=\frac{\sqrt{\varepsilon/\rho}}{K},\qquad
  L=\frac{L_d}{H}.
  \label{eq:params}
\end{equation}
The electric Rayleigh number $T$ represents the ratio of electric to viscous
forces and will be the stability parameter. The injection number $C$
characterizes the strength of charge injection. The mobility number $M$ is
the ratio of the hydrodynamic mobility to the ionic mobility
\citep{Castellanos1998} and depends only on the properties of the fluid.
Finally, $L$ is the aspect ratio of the layer.

The nondimensional base state is
\begin{equation}
  \begin{split}
  &q_b(y)=\left(1+\frac{2C}{E_i}\,y\right)^{-1/2},\\
  &E_b(y)=E_i\left(1+\frac{2C}{E_i}\,y\right)^{1/2},
  \end{split}
  \label{eq:base_nd}
\end{equation}
where the injector field $E_i$ is determined by
\begin{equation}
  \frac{E_i^2}{3C}\left[\left(1+\frac{2C}{E_i}\right)^{3/2}-1\right]=1 .
  \label{eq:Es}
\end{equation}
The base pressure balances the Coulomb force, $p_b(y)=CM^2E_i\,y$.

With the scales of Eq.~\eqref{eq:scales}, the drift velocity of the ions relative to
the fluid is $\bE$, and velocities are measured in the drift-velocity scale
$KV_0/H$. In weak injection $E_b\simeq1$ across the layer, so a unit velocity is
the local drift velocity. In strong injection this is not the case. For $C=10$ the
hydrostatic field grows from $E_i=0.11$ at the injector to $1.49$ at the
collector, close to the space-charge-limited profile $\tfrac32\sqrt{y}$, and the
local drift velocity varies accordingly. Its average across the layer, however,
is exactly one in any state, conductive or convective, since $\int_0^1E_y\,\dd
y$ is the applied voltage. A maximum fluid velocity $U_{\max}>1$ therefore means
that the flow is faster than the mean ionic drift, so that convection competes
with migration in transporting the charge. We use this comparison below as an
indicator, not as a local criterion. Since $\bE$ is irrotational, the
transport velocity $\bu+\bE$ can only have closed orbits where the flow
overcomes the local drift. In the core of the layer $E_b$ is close to one, so
this requires $U_{\max}$ of order one. The charge obeys
$\partial_tq+(\bu+\bE)\cdot\nabla q=-Cq^2$, so it decreases monotonically along
the orbits of $\bu+\bE$. In a steady state it must therefore vanish on closed
orbits, which are not fed from the injector. This produces the charge void
discussed in Sec.~\ref{sec:minseed}.

\subsection{Linear modal stability}
\label{sec:linear}

In the linear stability analysis, the governing equations are linearized
introducing infinitesimal perturbations. The
resulting linearized equations determine the growth or decay of these
perturbations, and the onset of instability is identified with a neutral
perturbation, that is, a perturbation that neither grows nor decays. The
perturbations are defined with respect to the base state,
\begin{equation}
\begin{aligned}
  &\bu'=\bu,\qquad             &q'=q-q_b(y), \\
  &\phi'=\phi-\phi_b(y),\qquad &p'=p-p_b(y),
  \label{eq:perturbations}
\end{aligned}
\end{equation}
and sought in the form
\begin{equation}
  (\bu',p',\phi',q')=(\hat{\bu},\hat p,\hat\phi,\hat q)\exp(\sigma t),
  \label{eq:modal}
\end{equation}
where $\sigma$ is the temporal growth rate. The base state is linearly stable
when all admissible perturbations have $\mathrm{Re}\,\sigma<0$. In the present problem
the principle of exchange of stabilities applies, as proved by
\citet{AttenMoreau1972} for weak injection and verified numerically for strong
injection (Sec.~\ref{sec:transit}), and the marginal state is obtained for
$\sigma=0$. With this condition the mobility
parameter $M$ drops out of the linear stability problem, and the threshold
depends only on the injection parameter, the aspect ratio and $T$.

For a layer of infinite horizontal extent, translational invariance allows
the perturbations to be decomposed into Fourier modes,
\begin{equation}
  (\bu',p',\phi',q')=\left(\hat{\bu}(y),\hat p(y),\hat\phi(y),\hat q(y)\right)
  \exp(\mathrm{i}kx+\sigma t).
  \label{eq:fourier}
\end{equation}
For each horizontal wavenumber $k$, the linearized equations define an
eigenvalue problem for $T$. The neutral curve $T=T(k)$ gives the value at
which a perturbation of wavenumber $k$ becomes marginally unstable, and the
classical linear threshold is
\begin{equation}
  T_c=\min_k T(k),\qquad k_c=\arg\min_k T(k).
  \label{eq:Tc}
\end{equation}
This is essentially the procedure followed in the classical analysis of
\citet{AttenMoreau1972}. The instability results from the coupling between the
convective displacement of the space charge and the Coulomb force generated
by the resulting perturbation of the electric field. The critical value
depends strongly on the injection strength $C$. In particular, Atten and
Moreau obtained the well-known linear thresholds for the weak and strong
injection limits.

\citet{Perez2014} extended this formulation by solving the linearized partial
differential equations directly with finite elements. This becomes necessary
when rigid lateral walls are introduced, because the perturbations can no
longer be separated into independent functions of $x$ and $y$. After spatial
discretization, the marginal stability equations take the generalized
eigenvalue form
\begin{equation}
  \mathsf{A}\bm{x}=T\,\mathsf{B}\bm{x},
  \label{eq:gevp}
\end{equation}
where the eigenvector $\bm{x}$ contains the discrete velocity, pressure,
electric potential and charge density perturbations. The calculation
therefore provides not only the critical value of $T$ but also the complete
spatial structure of the corresponding linear eigenmode.

For $C=10$ and symmetry lateral walls,
\citet{Perez2014} obtained $T_c=164.09$ for a single roll of width $L_c=0.614$,
i.e., one half of the critical wavelength, in agreement with $T_c\simeq164.1$
and $L_c\simeq0.614$ of \citet{AttenMoreau1972}.

In the present work all computations, modal and non-modal, are carried out in a
laterally periodic layer of one critical wavelength, $L=2L_c=1.228$, which
contains two counter-rotating rolls. The marginal problem, Eq.~\eqref{eq:gevp}, is
assembled from the same discretization used for the optimization
(Sec.~\ref{sec:discretization}). The operator is the exact linearization, at the
base state, of the discrete residual of the time stepper, including the SUPG
stabilization. Its smallest eigenvalue $T_c^h$ is therefore the threshold of the
semi-discrete problem solved by the optimization, to which the non-modal
threshold of Sec.~\ref{sec:threshold} must converge.

The problem in the periodic layer is invariant under horizontal translations and
under the reflection $(x,u)\mapsto(-x,-u)$. Because the fields are $L$-periodic,
$x$ can be regarded as an angle, $\theta=2\pi x/L$, so that a translation
$x\mapsto x+a$ is a rotation of this circle by $2\pi a/L$. With the reflection,
these rotations form the group O(2). On a mode of wavenumber $k$,
$\bz'=A\,\hat{\bz}(y)\,e^{\mathrm ikx}+\text{c.c.}$, a translation multiplies
the complex amplitude by $e^{\mathrm ika}$, i.e., it rotates the pair
$(\mathrm{Re}\,A,\mathrm{Im}\,A)$, while the reflection maps $A$ to
$\bar A$. Since every direction of this plane can be rotated into any other, the
eigenvalue of each mode with $k\neq0$ is double, and the critical eigenspace
consists of rolls of arbitrary horizontal phase
\citep{Golubitsky1988,CrawfordKnobloch1991}. Due to the reflection symmetry and
the exchange of stabilities \citep{AttenMoreau1972}, they are steady. On a
uniform periodic mesh, such as the one used in the present work, the symmetry
group reduces to the dihedral group generated by translations by whole cells and
the reflection, which still leaves the critical eigenvalue exactly double.

Table~\ref{tab:modal} lists $T_c^h$ on four meshes. The threshold converges
monotonically with observed order $p=2.10$ to $T_\infty=164.06$, within $0.02\%$
of the published value. The production mesh of the rest of the paper, with
$62\times100$ cells per roll, gives $T_c^h=163.82$, $0.14\%$ below $T_\infty$.
As a cross-check, the discretization of \citet{Perez2014} (triangles, $P_2$
charge, no stabilization, $h=0.01$) gives $T_c=163.92$ in the periodic layer,
$0.1\%$ below the published value. 

\begin{table}
  \centering
  \caption{Discrete modal threshold $T_c^h$ in the periodic layer
    ($C=10$, $L=1.228$) for the discretization of
    Sec.~\ref{sec:discretization}.}
  \label{tab:modal}
  \begin{tabular}{lccc}
    \toprule
    Cells per roll & Cells in the layer & Unknowns & $T_c^h$ \\
    \midrule
    $31\times50$  & $62\times50$   & 43\,896  & 163.024 \\
    $47\times75$  & $94\times75$   & 99\,452  & 163.625 \\
    $62\times100$ & $124\times100$ & 174\,592 & 163.822 \\
    $93\times150$ & $186\times150$ & 392\,088 & 163.958 \\
    \midrule
    $h\to0$ ($p=2.10$) & & & 164.06 \\
    \bottomrule
  \end{tabular}
\end{table}

\begin{figure*}
  \centering
  \includegraphics[width=\linewidth]{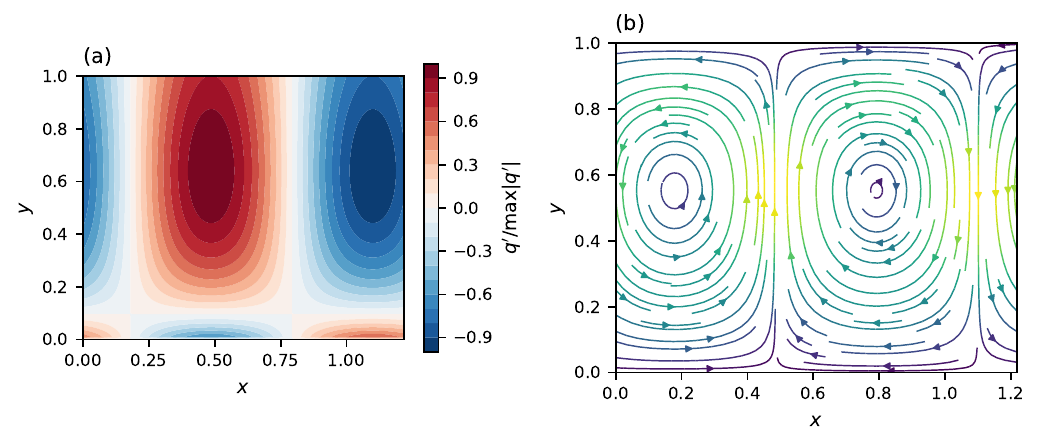}
  \caption{Critical eigenmode in the periodic layer ($C=10$, $L=1.228$,
    $62\times100$ cells per roll, $T_c^h=163.82$). (a) Charge density
    perturbation. (b) Streamlines of the velocity perturbation, colored by the
    speed. The horizontal phase is arbitrary.}
  \label{fig:eigenfunction}
\end{figure*}

Figure~\ref{fig:eigenfunction} shows the critical mode on the production mesh.
The charge perturbation consists of vertical columns of alternating sign
spanning the whole layer, with maxima at $y\simeq0.6$ and thin layers of
opposite sign next to the injector, and the velocity field consists of two
counter-rotating rolls. The charge perturbation is antisymmetric about the
center of each roll.

\section{Non-modal nonlinear stability analysis}
\label{sec:optimization}

\subsection{Optimization problem}
\label{sec:optproblem}

The aim of the nonlinear non-modal analysis is to identify the initial
perturbation that maximizes a given objective functional at a prescribed target
time $\tau$, which we will refer to as the horizon. The optimization problem is
defined by the governing equations, the initial perturbation amplitude and the
objective functional.

We consider initial perturbations of the charge only, with the velocity
initially at rest. This choice models fluctuations of the injected charge,
which are the natural disturbances of the hydrostatic state. The initial
amplitude is prescribed through
\begin{equation}
  E_0=\int_\Omega (q'_0)^2\dd\Omega .
  \label{eq:E0}
\end{equation}
Here $\Omega$ is the computational domain, with boundary
$\partial\Omega=\Gamma_i\cup\Gamma_c\cup\Gamma_s$, where $\Gamma_i$ and
$\Gamma_c$ denote the injecting and collecting electrodes and $\Gamma_s$ the
lateral (periodic) boundaries. The control $q'_0$ is the perturbation with respect
to the base state $q_b(y)$. Since the base state is hydrostatic, we choose
the total kinetic energy at the target time as the objective functional,
\begin{equation}
  J(\tau)=\frac12\int_\Omega|\bu(\tau)|^2\dd\Omega ,
  \label{eq:J}
\end{equation}
and measure the amplification of the initial perturbation by the gain
\begin{equation}
  G_q(\tau)=\frac{J(\tau)}{E_0}.
  \label{eq:Gq}
\end{equation}
Since $G_q$ compares a kinetic energy with a charge norm, its value depends on
the scales of Eq.~\eqref{eq:scales} and on the parameters $C$ and $M$. It is
therefore not directly comparable with the energy gains of
\citet{Zhang2015}, whose norm combines velocity and charge perturbations.
The optimization problem is
\begin{equation}
  \max_{q'_0} J(\tau)\quad\text{subject to}\quad
  \int_\Omega (q'_0)^2\dd\Omega=E_0 .
  \label{eq:optim}
\end{equation}

To obtain the gradient of the objective functional with respect to the initial
charge perturbation, we adopt a Lagrangian approach \citep{Kerswell2018}. The
governing equations are enforced through adjoint fields $\ua$, $\pa$, $\qa$ and
$\phia$, associated respectively with the momentum equation, incompressibility,
charge conservation and Poisson's equation. The augmented Lagrangian is
\begin{equation}
  \begin{aligned}
  &\mathcal{L}= J-\lambda\left(\int_\Omega (q'_0)^2\dd\Omega-E_0\right) \\
  &-\int_0^\tau\!\!\int_\Omega \ua\cdot\left[\frac{\partial\bu}{\partial t}
    +(\bu\cdot\nabla)\bu+\nabla p\right.\\
  &\qquad\qquad\left.-\frac{M^2}{T}\nabla^2\bu-CM^2q\bE\right] \dd\Omega\dd t \\
    &-\int_0^\tau\!\!\int_\Omega \pa\,\nabla\cdot\bu\dd\Omega\dd t\\
    &-\int_0^\tau\!\!\int_\Omega \qa\left[\frac{\partial q}{\partial t}+\nabla\cdot\left(q(\bu+\bE)\right)\right]\dd\Omega\dd t\\
    & -\int_0^\tau\!\!\int_\Omega \phia\left(\nabla^2\phi+Cq\right)\dd\Omega\dd t,
  \end{aligned}
  \label{eq:lagrangian}
\end{equation}
where $\lambda$ is the Lagrange multiplier associated with the amplitude
constraint. The adjoint equations and boundary conditions follow by requiring
the first variation of $\mathcal{L}$ to vanish for arbitrary admissible
variations of the direct variables $\bu$, $p$, $q$ and $\phi$. The detailed
derivation is given in Appendix~\ref{app:adjoint}, together with the final form
of the adjoint equations and boundary conditions.

\subsection{Optimization algorithm}
\label{sec:algorithm}

The nonlinear optimization is carried out by direct--adjoint iteration. Starting
from an admissible initial charge perturbation $q'_0$ satisfying Eq.~\eqref{eq:E0},
the nonlinear direct problem is integrated forward in time from $t=0$ to
$t=\tau$ and the objective functional, Eq.~\eqref{eq:J}, is evaluated. The adjoint
equations are then integrated backward in time from $t=\tau$ to $t=0$ to obtain
the gradient of $J$ with respect to the initial perturbation,
\begin{equation}
  g=\frac{\delta J}{\delta q'_0}.
  \label{eq:grad}
\end{equation}
Here $g$ is the $L^2$ Riesz representer of the derivative of $J$
\citep{Brezis2011}. The adjoint provides this derivative as a linear
functional of the variation of $q'_0$, and $g$ is the function that
represents it in the $L^2$ inner product, the same inner product that defines
the amplitude constraint, Eq.~\eqref{eq:E0}. The control space is
$\mathcal{A}=\{q'_0 : q'_0=0 \text{ on } \Gamma_i\}$, so that, after
discretization, $g$ is the solution of a mass-matrix problem with homogeneous
Dirichlet conditions on the injector. Using the mass matrix, rather than the
raw derivative with respect to the degrees of freedom, makes the gradient
independent of the mesh \citep{Schwedes2017}.

Because the control is constrained to the hypersphere defined by Eq.~\eqref{eq:E0}, only the
component of the gradient tangent to this manifold is admissible. The
projected gradient is
\begin{equation}
  g_T=g-\frac{\langle g,q'_0\rangle}{\langle q'_0,q'_0\rangle}\,q'_0
     =g-\frac{\langle g,q'_0\rangle}{E_0}\,q'_0 .
  \label{eq:proj}
\end{equation}
The control is updated along an ascent direction $d$ tangent to the sphere,
\begin{equation}
  \tilde q'_0=q'_0+\alpha d .
  \label{eq:update}
\end{equation}
In the simplest algorithm, the value of $d$ is computed using the steepest
ascent $d=g_T$. In this work, $d$ is computed with the Riemannian limited-memory BFGS
method \citep{RingWirth2012,Absil2008}, and the step length $\alpha$ is selected
by an Armijo backtracking line search \citep{Armijo1966,Nocedal2006}. The
updated perturbation is projected back onto the constraint,
\begin{equation}
  q'_0\leftarrow\sqrt{\frac{E_0}{\int_\Omega(\tilde q'_0)^2\dd\Omega}}\;\tilde q'_0 .
  \label{eq:retraction}
\end{equation}
The forward integration, backward adjoint integration, gradient projection,
line search and renormalization are repeated until the projected gradient is
sufficiently small. Convergence is monitored through the ratio
\begin{equation}
  \frac{\|g_T\|}{\|g\|},
  \label{eq:ratio}
\end{equation}
which vanishes at a constrained stationary point. Convergence is declared
when $\|g_T\|/\|g\|\le10^{-4}$. Close to an optimum the error in $J$ is
quadratic in this ratio, so $G_q$ is then accurate to better than $10^{-7}$.
The adjoint gradient was verified with Taylor tests at the analytical initial guess and
at converged optima, in the linear regime and at finite amplitude. In all
cases the Taylor remainder converges at rate $2.00$, two adjoint solutions on
the same trajectory agree to $10^{-15}$, and the Riesz map reproduces the
discrete derivative to round-off.

\subsection{Discretization and discrete adjoint}
\label{sec:discretization}

The gradient, Eq.~\eqref{eq:grad}, is obtained with the discrete adjoint generated
automatically by the Python library Pyadjoint \citep{Mitusch2019} from the
Firedrake implementation of the forward problem \citep{Rathgeber2016}. The
choice of this automatic-differentiation framework motivates the monolithic
formulation of the direct problem.

It is usual to apply split methods to the time integration of incompressible
flows, due to the saddle-point structure of the equations. However, the adjoint
of a split scheme is not the split of the adjoint, and the adjoint of a
projection scheme is not a projection scheme.  Therefore, we adopt a monolithic
formulation of the direct problem. Equations~\eqref{eq:ns}--\eqref{eq:poisson}, together with the boundary conditions of
Eq.~\eqref{eq:bc}, are discretized in time as a single system, and the velocity,
pressure, electric potential and charge density are solved fully coupled at
each time step. 
The unknowns are the perturbation fields
$\bz'=(\bu',p',q',\phi')$ with respect to the base state of Eq.~\eqref{eq:base_nd}.
Writing $\mathcal{R}(\bz)$ for the discrete residual of the full equations,
Eqs.~\eqref{eq:ns}--\eqref{eq:poisson}, each time step solves
\begin{equation}
  \mathcal{R}'(\bz') \equiv \mathcal{R}(\bz_b+\bz')-\mathcal{R}(\bz_b)=0 ,
  \label{eq:residual_difference}
\end{equation}
where $\bz_b$ is the interpolated base state. Subtracting the discrete base
residual makes $\bz'=0$ an exact solution of the discrete problem. Without this
subtraction, the interpolation error of the base state would act as a spurious
forcing.

The domain is the periodic layer $[0,L)\times[0,1]$ with $L=1.228$, one
critical wavelength (two rolls), discretized with a structured
quadrilateral mesh. The mesh is invariant under horizontal translations by
one cell and under the reflection $x\mapsto-x$. We use $Q_2$ elements for
velocity and potential, and $Q_1$ elements for pressure and charge density.
Since diffusion is neglected, the charge transport equation, Eq.~\eqref{eq:charge},
is hyperbolic, and its discretization is stabilized with a
streamline-upwind Petrov--Galerkin (SUPG) term \citep{Brooks1982} with the
time-step-independent parameter
\begin{equation}
  \tau_{\rm SUPG}=\Big[\big(2|\bv_q|/h_K\big)^2+(2Cq)^2\Big]^{-1/2},
  \quad \bv_q=\bu-\nabla\phi ,
  \label{eq:tau_supg}
\end{equation}
where $h_K$ is the cell diameter. Time integration uses the second-order
backward differentiation formula (BDF2), started with one backward-Euler step
\citep{HairerWanner1996,AscherPetzold1998}. At each step the monolithic
nonlinear system is solved with Newton's method and a direct (LU) linear solver.
The pressure null space is removed in both the direct and the adjoint problems,
and null pivots of the LU factorizations are detected explicitly. Without this
safeguard, the singular adjoint system can occasionally return a corrupted
gradient. The complete weak formulation is given in Appendix~\ref{app:weak}.

During the forward integration, Pyadjoint records on a tape every operation that
contributes to $J$, namely the interpolation of the initial perturbation, each
nonlinear time step, and the assembly of the objective functional. Each time
step is recorded as a single block, defined by its residual in variational form,
and not by the Newton iterations used to solve it. In the reverse sweep the tape
is traversed backwards. For each time step, the Jacobian of the residual with
respect to the unknowns of that step is obtained by symbolic differentiation of
the variational form and its transpose is assembled. The adjoint variables of
the step are then the solution of a single linear system with this transposed
Jacobian, whose right-hand side collects the sensitivities propagated from the
later steps and from $J$. The backward integration is therefore linear, and each
adjoint step costs one linear solve, against the several Newton iterations of
the corresponding forward step. The result of the sweep is the derivative of the
discrete objective functional with respect to the degrees of freedom of $q'_0$,
a linear functional from which the gradient $g$ is obtained by the Riesz map of
Sec.~\ref{sec:algorithm}. 

Since the adjoint is derived from the discrete equations actually solved, it
contains the transposed SUPG contributions, which do not appear in the
continuous adjoint equations of Appendix~\ref{app:adjoint}. The optimization is
therefore internally consistent with the numerical forward problem, and
convergence towards the continuous optimization problem must be assessed through
spatial and temporal refinement.

\subsection{Amplitude measure, symmetry and computational configuration}
\label{sec:setup}

Amplitudes are given by the initial charge energy $E_0$ of Eq.~\eqref{eq:E0},
integrated over the layer ($|\Omega|=L=1.228$). For physical interpretation
we also quote the root-mean-square initial charge perturbation
\begin{equation}
  q'_{\rm rms}=\left(\frac{E_0}{|\Omega|}\right)^{1/2},
  \label{eq:qrms}
\end{equation}
which is expressed in units of the injected charge density $q_i$.

Equations~\eqref{eq:ns}--\eqref{eq:poisson}, the base state and the periodic
mesh are invariant under the reflection $\mathcal{P}:x\mapsto-x$ (modulo $L$),
under which $(u_x,u_y)\mapsto(-u_x,u_y)$ and $p$, $q$, $\phi$ are mapped to
their mirror images. The subspace of $\mathcal{P}$-even fields is therefore
invariant under the full nonlinear evolution. It is exactly the single roll with
symmetry walls of \citet{Perez2014}, which the periodic layer contains as a
special case. The $\mathcal{P}$-odd subspace is invariant only under the
linearized dynamics, because $q'\mapsto-q'$ is not a symmetry of the quadratic
terms in Eq.~\eqref{eq:charge}. Finite-amplitude optimizations are started from the
seed $y[\cos kx+\tfrac12\sin2kx]$, $k=2\pi/L$, which has no reflection axis and
lets the optimizer break both the reflection and the translation symmetries.

Unless stated otherwise, all results use $C=10$, $M=10$, the periodic layer
discretized with $62\times100$ cells per roll ($124\times100$ in the layer) and
$\Delta t=0.02$. Section~\ref{sec:validation} shows that this resolution
carries errors of about $0.02\%$ (spatial) and $0.2\%$ (temporal) on optimal
gains, and shifts the threshold to $T_c^h=163.82$. To keep every case at the
intended distance from threshold, values of the electric Rayleigh number are
referred to the discrete threshold. A case defined in terms of
$\Delta T=T_c-T$ is computed at $T=T_c^h-\Delta T$. In particular, the
reference subcritical case of the validation and of
Sec.~\ref{sec:nonlinear}, $\Delta T=4.1$ (i.e., $T=160$ for $T_c=164.09$), is
computed at $T=159.73$.

\section{Numerical validation}
\label{sec:validation}

\subsection{Spatial and temporal convergence}

The discretization was assessed with the small-amplitude optimization
($E_0=10^{-8}$, $\tau=0.4$). Table~\ref{tab:spatial} shows the optimal gain
at fixed $T=158.93$ on three successively doubled meshes at $\Delta t=0.02$.
The gain converges monotonically with observed order about $2.5$. The
production mesh ($62\times100$ cells per roll) differs from the finest one by
$0.015\%$, and the coarsest one ($31\times50$) by $0.10\%$.

\begin{table}
  \centering
  \caption{Spatial convergence of the optimal gain ($T=158.93$,
    $E_0=10^{-8}$, $\tau=0.4$, $\Delta t=0.02$).}
  \label{tab:spatial}
  \begin{tabular}{lcc}
    \toprule
    Cells per roll & $G_q$ & Relative difference to finest (\%) \\
    \midrule
    $31\times50$   & 629.564 & 0.099 \\
    $62\times100$  & 629.037 & 0.015 \\
    $124\times200$ & 628.941 & 0 \\
    \bottomrule
  \end{tabular}
\end{table}

Table~\ref{tab:temporal} shows the temporal convergence on the production
mesh at the reference value $T=159.73$. The differences decrease by a factor
of four per halving of $\Delta t$, and the observed order is $2.09$, as expected
for BDF2. Richardson extrapolation gives $G_q=641.652$. At the working
time step $\Delta t=0.02$ the temporal error is $0.18\%$, which is the
dominant discretization error of the production computations.

\begin{table}
  \centering
  \caption{Temporal convergence of the optimal gain on the production mesh
    ($T=159.73$, $E_0=10^{-8}$, $\tau=0.4$).}
  \label{tab:temporal}
  \begin{tabular}{lrcc}
    \toprule
    $\Delta t$ & $N_t$ & $G_q$ & Relative difference to finest (\%) \\
    \midrule
    $2.0\times10^{-2}$  & 20  & 640.503 & 0.179 \\
    $1.0\times10^{-2}$  & 40  & 641.403 & 0.038 \\
    $5.0\times10^{-3}$  & 80  & 641.596 & 0.008 \\
    $2.5\times10^{-3}$  & 160 & 641.639 & 0.002 \\
    $1.25\times10^{-3}$ & 320 & 641.649 & 0 \\
    \bottomrule
  \end{tabular}
\end{table}

\subsection{Determination of the linear-amplitude limit and recovery of the linear instability threshold}
\label{sec:threshold}

\begin{figure}
  \centering
  \includegraphics[width=\columnwidth]{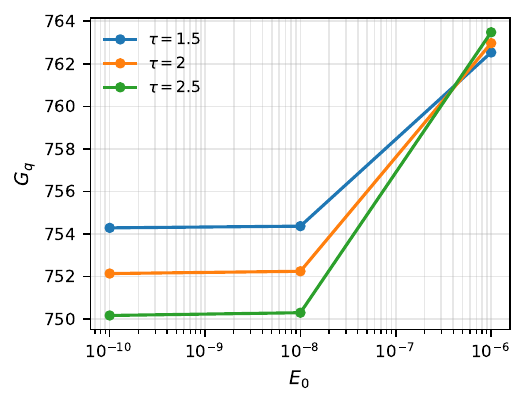}
  \caption{Optimal gain as a function of the initial charge energy
    $E_0$ near threshold ($T=163.73$) for three horizons.}
  \label{fig:linear_limit}
\end{figure}

In the linearized problem the velocity at time $\tau$ is proportional to $q'_0$,
so $J$ and $E_0$ are both quadratic in the amplitude of $q'_0$ and the gain
becomes independent of $E_0$ as $E_0\to0$. Table~\ref{tab:linear_limit} and
Fig.~\ref{fig:linear_limit} show the optimal gain near threshold ($T=163.73$,
$\Delta T=0.1$) for $\tau=1.5$, $2$, $2.5$ and three amplitudes. Between
$E_0=10^{-10}$ and $10^{-8}$ the gain changes by at most $0.017\%$, whereas
$E_0=10^{-6}$ already deviates by $1.1$--$1.8\%$, increasing with $\tau$. We
therefore use $E_0=10^{-8}$ to represent the linear regime.

\begin{table}
  \centering
  \caption{Optimal gain versus initial amplitude near threshold
    ($T=163.73$, production mesh).}
  \label{tab:linear_limit}
  \begin{tabular}{ccccc}
    \toprule
    $\tau$ & $G_q(10^{-10})$ & $G_q(10^{-8})$ & $G_q(10^{-6})$
      & $\frac{G_q(10^{-8})-G_q(10^{-10})}{G_q(10^{-10})}$ (\%) \\
    \midrule
    1.5 & 754.283 & 754.364 & 762.536 & 0.011 \\
    2.0 & 752.134 & 752.240 & 762.973 & 0.014 \\
    2.5 & 750.162 & 750.292 & 763.485 & 0.017 \\
    \bottomrule
  \end{tabular}
\end{table}

\begin{figure*}
  \centering
  \includegraphics[width=\linewidth]{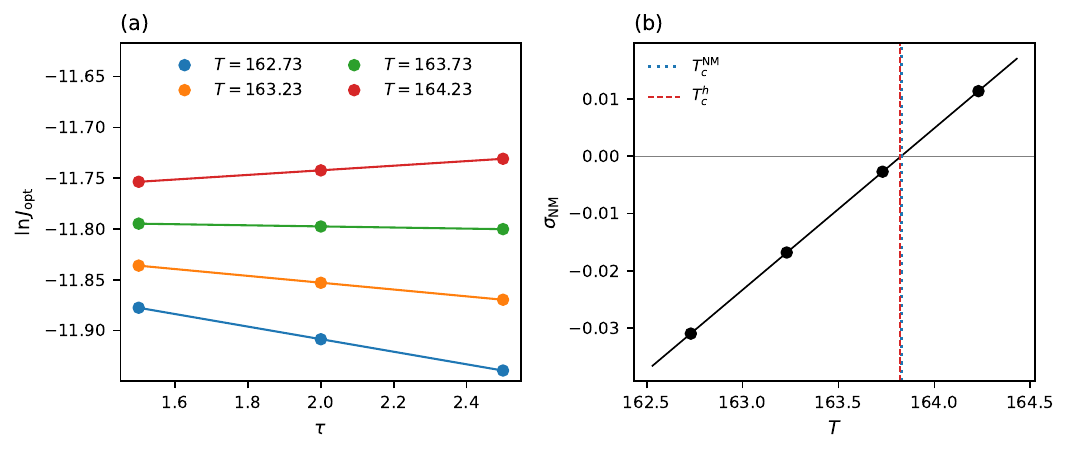}
  \caption{Recovery of the linear threshold. (a) $\ln( J_{\rm opt})$ versus
    $\tau$ with least-squares fits ($E_0=10^{-8}$). (b) $\sigma_{\rm NM}(T)$.
    Dotted: zero crossing $T_c^{\rm NM}=163.827$. Dashed: discrete modal
    threshold $T_c^h=163.822$.}
  \label{fig:threshold}
\end{figure*}

For small amplitude and long horizon the optimal response is dominated by
the least stable eigenmode, so that
\begin{equation}
  J_{\rm opt}(\tau)\simeq A(T)\exp\left[2\sigma(T)\tau\right].
  \label{eq:asymptotic}
\end{equation}
The asymptotic growth rate is obtained from a least-squares fit of
$\ln(J_{\rm opt})=a+b\tau$ over $\tau\in\{1.5,2,2.5\}$, and
$\sigma_{\rm NM}=b/2$. The fits for $E_0=10^{-8}$ and four values of $T$
around the discrete threshold are shown in Fig.~\ref{fig:threshold}(a).
All coefficients of determination exceed $0.9994$. The growth rate is linear
in $T$ with slope $\mathrm{d}\sigma_{\rm NM}/\mathrm{d}T=0.0282$
[Fig.~\ref{fig:threshold}(b) and Table~\ref{tab:threshold}], and its zero
gives the non-modal threshold
\begin{equation}
  T_c^{\rm NM}=163.827 ,
\end{equation}
which coincides with the discrete modal threshold $T_c^h=163.822$ of the same
mesh to within $0.003\%$. The direct--adjoint
optimization thus recovers the modal threshold of its own discretization
essentially exactly, and the mesh convergence of $T_c^h$
(Table~\ref{tab:modal}) carries it to the
continuous value.

\begin{table}
  \centering
  \caption{Asymptotic growth rates on the production mesh ($E_0=10^{-8}$,
    $\Delta t=0.02$), fitted over $\tau\in\{1.5,2,2.5\}$.}
  \label{tab:threshold}
  \begin{tabular}{ccc}
    \toprule
    $T$ & $\sigma_{\rm NM}$ & $R^2$ \\
    \midrule
    162.73 & $-3.092\times10^{-2}$ & 0.999995 \\
    163.23 & $-1.679\times10^{-2}$ & 0.999985 \\
    163.73 & $-2.707\times10^{-3}$ & 0.999419 \\
    164.23 & $+1.133\times10^{-2}$ & 0.999967 \\
    \midrule
    \multicolumn{3}{c}{$T_c^{\rm NM}=163.827$, \quad $T_c^h=163.822$} \\
    \bottomrule
  \end{tabular}
\end{table}

\section{Nonlinear optimal perturbations}
\label{sec:nonlinear}

The finite-amplitude calculations are performed at $T=159.73$
($\Delta T=4.1$, linearly stable) for the horizons
$\tau\in\{0.1,0.5,1,1.5,2\}$. For each horizon the optimization is continued
in amplitude over $E_0=10^{-8},10^{-7},\dots,10^{-4}$,
each stage starting from the
optimum of the previous one and the first from the symmetry-breaking seed of
Sec.~\ref{sec:setup}. All optimizations converged to
$\|g_T\|/\|g\|<10^{-4}$.

\subsection{Finite-amplitude optimal growth}

\begin{figure*}
  \centering
  \includegraphics[width=\linewidth]{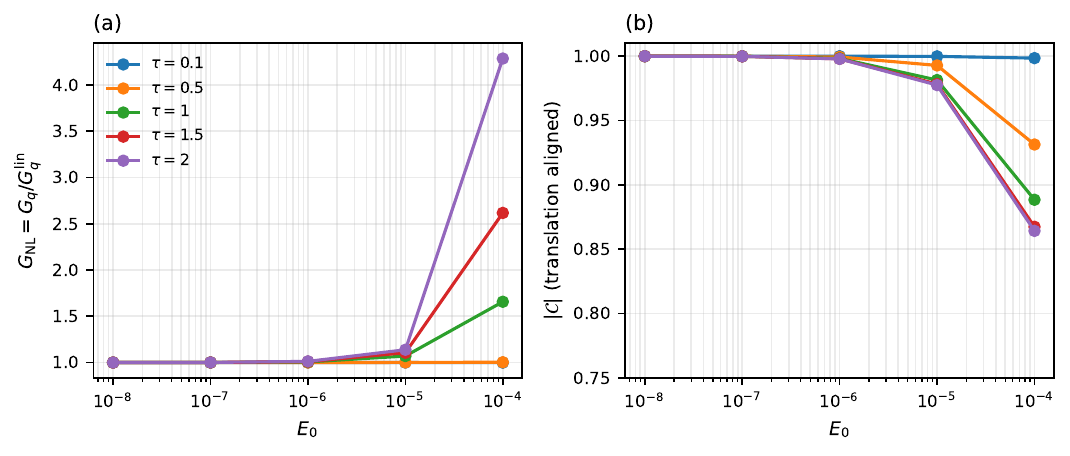}
  \caption{Finite-amplitude optimal growth at $T=159.73$. (a) Optimal gain
    normalized by its small-amplitude value. (b) Translation-aligned
    correlation of Eq.~\eqref{eq:corr} between the optimal perturbation at
    amplitude $E_0$ and the small-amplitude optimum.}
  \label{fig:amplitude}
\end{figure*}

Figure~\ref{fig:amplitude} shows the optimal gain $G_q=J_{\rm opt}/E_0$
normalized by its small-amplitude value, $G_{\rm NL}=G_q/G_q^{\rm lin}$, and the
structural similarity defined below. In the linear regime the optimal gain
is not monotonic in $\tau$. It takes the values $G_q^{\rm lin}=368.7$, $684.6$, $601.2$, $539.9$
and $480.3$ for $\tau=0.1$, $0.5$, $1$, $1.5$ and $2$. Transient
amplification peaks near $\tau\approx0.5$, after which the response decays at
the asymptotic rate $\sigma_{\rm NM}<0$ of this linearly stable state.

Up to $E_0\simeq10^{-6}$ the gain is independent of the amplitude to
within $1.3\%$ for all horizons. At larger amplitude the horizons separate.
At $E_0=10^{-5}$, $G_{\rm NL}=1.07$, $1.11$ and $1.14$ for $\tau=1$,
$1.5$ and $2$, and at $E_0=10^{-4}$, $G_{\rm NL}=1.66$, $2.62$ and $4.29$.
For the shortest horizons the nonlinearity has no time to act, with
$G_{\rm NL}=0.999$ ($\tau=0.1$) and $1.003$ ($\tau=0.5$) at the largest
amplitude. At $E_0=10^{-4}$ and $\tau=2$ the maximum velocity
reaches $U_{\max}=1.10$ at $t=\tau$, i.e., of the order of the drift-velocity scale. The
optimal perturbation is driving the layer towards the convective state
within the horizon.

\subsection{Structure of the finite-amplitude optimal perturbation}

In the periodic layer every translate of an optimum is also an optimum, so
the similarity between optima at different amplitudes is measured by the
correlation maximized over translations,
\begin{equation}
  |\mathcal{C}(E_0,\tau)|=\max_{s\in[0,L)}
  \frac{\big|\langle q'_0(x+s,y;E_0,\tau),\,q'_0(x,y;E_{0,\rm ref},\tau)\rangle\big|}
       {\|q'_0(E_0,\tau)\|\,\|q'_0(E_{0,\rm ref},\tau)\|},
  \label{eq:corr}
\end{equation}
with $E_{0,\rm ref}=10^{-8}$.
The correlation stays above $0.997$ up to $E_0=10^{-6}$ and decreases
to $0.93$, $0.89$, $0.87$ and $0.86$ for $\tau=0.5$, $1$, $1.5$ and $2$ at
$E_0=10^{-4}$ [Fig.~\ref{fig:amplitude}(b)]. This means that the structure of
the optimal perturbation changes with the amplitude of the initial perturbation.

\begin{figure*}
  \centering
  \includegraphics[width=\linewidth]{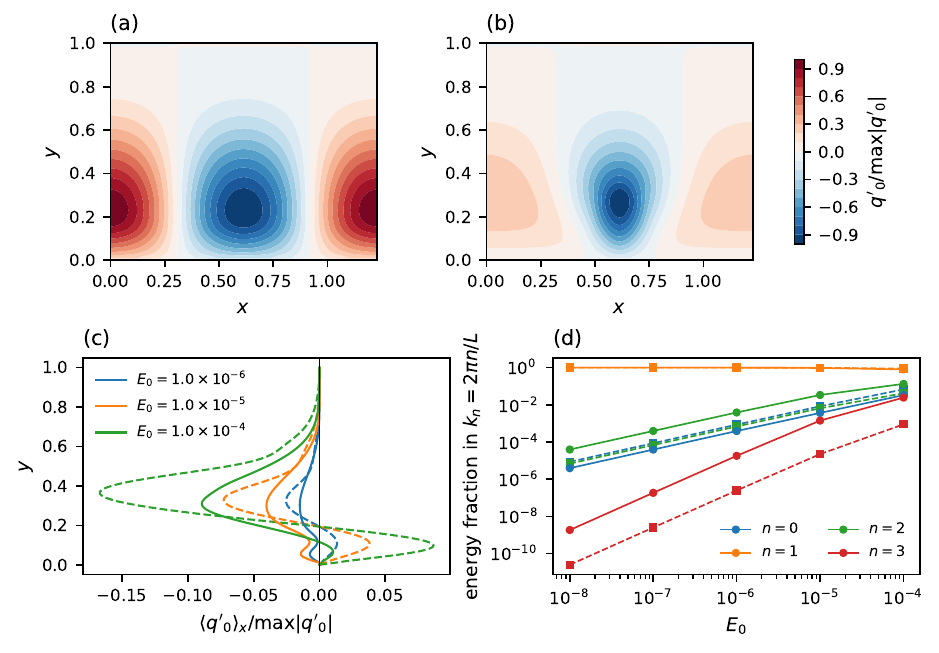}
  \caption{Optimal initial charge perturbation at $T=159.73$. (a)
    Small-amplitude optimum, $E_0=10^{-8}$, $\tau=1$. (b)
    Finite-amplitude optimum, $E_0=10^{-4}$, $\tau=1$. Both are
    normalized by $\max|q'_0|$. (c) Horizontal mean
    $\langle q'_0\rangle_x(y)/\max|q'_0|$ for the three largest amplitudes,
    for $\tau=1$ (solid) and $\tau=0.5$ (dashed). (d) Fraction of the
    perturbation energy in the horizontal wavenumbers $k_n=2\pi n/L$, with the
    same line styles.}
  \label{fig:structure}
\end{figure*}

Figure~\ref{fig:structure} shows this change for $\tau=1$. The small-amplitude
optimum ($E_0=10^{-8}$) is a single horizontal Fourier mode concentrated near the
injector. The maximum is at $y\simeq0.25$, much lower than the critical
eigenmode (Fig.~\ref{fig:eigenfunction}), with lobes of equal strength ($\max
q'_0=2.17\times10^{-4}$, $\min q'_0=-2.20\times10^{-4}$). For $E_0=10^{-4}$ the
optimum is strongly asymmetric in sign. It consists of a compact charge deficit ($\min
q'_0=-3.95\times10^{-2}$) flanked by weak excess regions ($\max
q'_0=1.10\times10^{-2}$). The asymmetry appears in the horizontal Fourier
content as a horizontally uniform component $k_0=0$, a charge deficit in the
bulk of the layer, and a second harmonic $k_2=2k$
[Figs.~\ref{fig:structure}(c) and \ref{fig:structure}(d)]. Their energy fractions grow linearly with
$E_0$, as expected for quadratic interactions of the fundamental mode, and reach
$3.3\%$ ($k_0$) and $13.4\%$ ($k_2$) of the perturbation energy for $\tau=1$ at
$E_0=10^{-4}$ ($6.7\%$ and $4.2\%$ for $\tau=0.5$).
The third harmonic $k_3=3k$, produced by the interaction of the fundamental
with $k_2$, is of third order in the amplitude, and its energy fraction grows
as $E_0^2$, from $1.8\times10^{-9}$ at $E_0=10^{-8}$ to $2.4\%$ at
$E_0=10^{-4}$ for $\tau=1$. Between $E_0=10^{-5}$ and $10^{-4}$ all fractions
depart from these power laws (the second harmonic grows by a factor of $4$
instead of $10$), and the fundamental retains only $80\%$ of the perturbation
energy. At this amplitude the optimal trajectory reaches $U_{\max}=0.80$ within
the horizon, close to the drift-velocity scale, and higher-order interactions
are no longer small.

Although the seed and the optimizer allow it, no optimum breaks the
reflection symmetry. The energy fraction odd about the best reflection axis
stays below $5\times10^{-8}$ for all amplitudes and horizons. A
single-stage optimization at $E_0=10^{-4}$, $\tau=1$, started
directly from the symmetry-breaking seed (without continuation) converges to
$G_q=996.117$, the value obtained by continuation, with
$\|g_T\|/\|g\|=7\times10^{-6}$. The finite-amplitude optimum is therefore
mirror-symmetric and coincides with the optimum of a single roll with
symmetry walls at the same amplitude per roll.

\subsection{Nonlinear mechanism: asymmetry between charge excess and deficit}
\label{sec:mechanism}

The sign asymmetry can be traced to the nonlinear part of the charge equation.
Using Eqs.~\eqref{eq:poisson} and \eqref{eq:div}, Eq.~\eqref{eq:charge} reads
$\partial_tq+(\bu+\bE)\cdot\nabla q+Cq^2=0$. Inserting $q=q_b+q'$ and
$\bE=\bE_b+\bE'$ and subtracting the base state gives
\begin{equation}
  \partial_t q'+(\bu+\bE_b+\bE')\cdot\nabla q'+(\bu+\bE')\cdot\nabla q_b
  =-\,C\,(2q_b+q')\,q' .
  \label{eq:charge_perturbation}
\end{equation}
The right-hand side is the Coulomb self-repulsion of the perturbation
charge. Its local relaxation rate $C(2q_b+q')$ is larger for a charge excess
($q'>0$) than for a charge deficit ($q'<0$) of the same magnitude. In the
linearized problem it is $2Cq_b$ for both signs, which is why the linear
optimum has symmetric lobes. At finite amplitude a deficit relaxes more
slowly and persists longer to modify the Coulomb force that drives the roll.

The interpretation can be tested by reversing the sign of an optimum, which
leaves its spatial structure and amplitude unchanged but turns the charge
deficit into an excess. Just above the critical amplitude
($E_0=3.492\times10^{-5}$, $0.4\%$ above $E_0^c$, Sec.~\ref{sec:minseed}) the
optimum reaches the convective state, whereas its sign reversal decays to
rest without approaching the edge state [Fig.~\ref{fig:modal_seed}(a)].
Section~\ref{sec:modal} shows that the two signs separate within the first time
unit.

\subsection{Minimal seed}
\label{sec:minseed}

For values of $T$ between $T_{\rm NL}$ and the linear threshold, both the
hydrostatic and the convective states are stable (Fig.~\ref{fig:sketch}).
The minimal seed is the smallest initial perturbation that, starting from the
hydrostatic state, reaches the convective state. Its amplitude
$E_0^c(T)$ measures how robust the hydrostatic state is. Following
\citet{PringleKerswell2010} it is approached from below with optimal
perturbations. For each amplitude, the perturbation maximizing $J(\tau)$ with
$\tau=1$ is computed and evolved to $t_{\rm end}=20$ ($40$ or $60$ where needed,
see below). The amplitude is scanned by decades and then bracketed by bisection
in $\log E_0$, each step restarted from the nearest non-transitioning optimum.
A trajectory is classified as transitioning when, at $t_{\rm end}$, the
maximum fluid velocity exceeds the drift-velocity scale, $U_{\max}\ge1$, as on
the convective branch \citep{Wu2015}, and its kinetic energy has reached a
plateau, $K(t_{\rm end})\ge0.97\,K(0.75\,t_{\rm end})$. All the other
trajectories decay, with $U_{\max}(t_{\rm end})<0.5$ and decreasing. The
convective states reached have $U_{\max}\ge1.34$ (Table~\ref{tab:minseed}),
well above the threshold, so the classification does not depend on its exact
value. A transitioning optimum proves
$E_0\ge E_0^c$ irrespective of how well it is converged. A decaying optimum is
accepted as a lower bound only if its optimization reached
$\|g_T\|/\|g\|\le10^{-2}$. The searches use this tolerance and up to 12
iterations, increased to 30 iterations and $10^{-3}$ for $\Delta T\ge46.6$,
where the optimization at large amplitude is harder. Trajectories are
integrated to $t=40$ at $\Delta T=4.1$ and $46.6$--$49.1$, and to $t=60$ at
$\Delta T=51.6$ and $54.1$.

\begin{figure*}
  \centering
  \includegraphics[width=\linewidth]{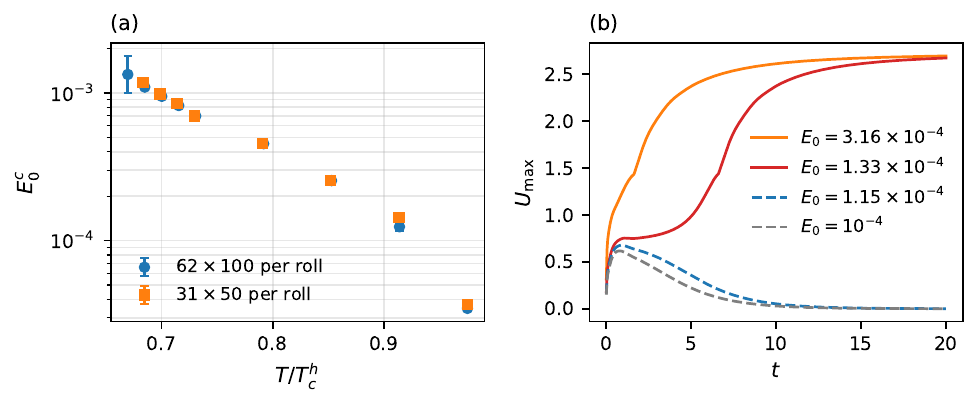}
  \caption{Minimal seed. (a) Critical amplitude $E_0^c$ versus $T/T_c^h$ on
    the production mesh and on a mesh with half the resolution ($31\times50$
    cells per roll), at the same distance from the discrete threshold of each
    mesh. (b) Maximum velocity of the
    optimal trajectories at $T=149.73$ ($\Delta T=14.1$) for four amplitudes of
    the search, namely the two that bracket $E_0^c$ and one further amplitude on each
    side. Solid lines are transitioning trajectories and dashed lines non-transitioning ones.}
  \label{fig:minseed}
\end{figure*}

The classification is unambiguous away from the bracket. Decaying
trajectories reach $K(t_{\rm end})\lesssim10^{-11}$ in most cases, and
transitioning ones saturate on the convective state
[Fig.~\ref{fig:minseed}(b)]. Close to the critical amplitude, trajectories
approach a state of intermediate amplitude and remain near it for several
time units before either decaying or departing towards convection. At
$T=149.73$ the smallest transitioning optimum stays at $U_{\max}\simeq0.75$
until $t\approx3$ and crosses $U_{\max}=1$ only at $t=5.1$.

\begin{table}
  \centering
  \caption{Minimal-seed amplitude bracket $E_0^{\rm lo}<E_0^c\le E_0^{\rm hi}$
    ($\tau=1$) at a distance $\Delta T=T_c^h-T$ from the discrete threshold
    $T_c^h=163.82$. $U_{\max}^{\rm conv}$ is the maximum velocity of the convective
    state reached.}
  \label{tab:minseed}
  \small
  \begin{tabular}{ccccc}
    \toprule
    $\Delta T$ & $T$ & $T/T_c^h$ & $E_0^c$ & $U_{\max}^{\rm conv}$ \\
    \midrule
    4.1  & 159.73 & 0.975 & $(3.461, 3.492]\times10^{-5}$ & 2.97 \\
    14.1 & 149.73 & 0.914 & $(1.15, 1.33]\times10^{-4}$ & 2.69 \\
    24.1 & 139.73 & 0.853 & $(2.37, 2.74]\times10^{-4}$ & 2.39 \\
    34.1 & 129.73 & 0.792 & $(4.22, 4.87]\times10^{-4}$ & 2.06 \\
    44.1 & 119.73 & 0.731 & $(6.49, 7.50]\times10^{-4}$ & 1.70 \\
    46.6 & 117.23 & 0.716 & $(8.06, 8.35]\times10^{-4}$ & 1.61 \\
    49.1 & 114.73 & 0.700 & $(9.31, 9.65]\times10^{-4}$ & 1.50 \\
    51.6 & 112.23 & 0.685 & $(1.075, 1.114]\times10^{-3}$ & 1.40 \\
    54.1 & 109.73 & 0.670 & $(1.00, 1.78]\times10^{-3}$ & 1.34 \\
    \bottomrule
  \end{tabular}
\end{table}

Table~\ref{tab:minseed} and Fig.~\ref{fig:minseed} give the result. The
minimal seed grows monotonically as the layer moves away from threshold.
Between $\Delta T=14.1$ and $44.1$ the critical amplitude increases by a
factor of about $5.6$, roughly as $E_0^c\propto\Delta T^{1.5}$, and by a
further factor of $1.6$ between $\Delta T=44.1$ and $51.6$. The
root-mean-square perturbation, Eq.~\eqref{eq:qrms}, at $E_0^c$ grows from $0.53\%$
of the injected charge at $\Delta T=4.1$ to $2.5\%$ at $\Delta T=44.1$ and
$3.0\%$ at $\Delta T=51.6$. At the same time the convective state reached
weakens steadily, $U_{\max}^{\rm conv}$ decreasing from $3.0$ to $1.3$ and its
kinetic energy from $1.10$ to $0.32$ (Table~\ref{tab:minseed}).

The minimal seed is converged in resolution to the precision of the
bisection. On a mesh with half the resolution in each direction ($31\times50$
cells per roll), the brackets are identical or higher by one bisection step
[Fig.~\ref{fig:minseed}(a)].

\paragraph{Lower end of the subcritical range.}
The convective state reached by the transitioning trajectories weakens
steadily as $T$ decreases (Table~\ref{tab:minseed}), which indicates that the
convective branch ends at a nonlinear threshold $T_{\rm NL}$, at a
saddle-node bifurcation where it merges with the branch of edge states
[Fig.~\ref{fig:sketch}(a)]. A convective state still exists at the lowest
value of the search, $T/T_c^h=0.670$ ($\Delta T=54.1$), with $U_{\max}=1.34$,
already close to the drift-velocity scale, and $K=0.32$. The end of the branch thus
lies just below,
\begin{equation}
  T_{\rm NL}\approx0.67\,T_c .
\end{equation}
Down to the end of the
branch the minimal seed is a perturbation of a few percent of the injected
charge. Below $T_{\rm NL}$ transition becomes impossible because the
convective branch itself ceases to exist.

This value agrees with direct numerical simulations of the convective branch. In
a cell of width $0.614$ with stress-free, zero-flux lateral boundaries, which is
equivalent to the $\mathcal{P}$-even subspace of our periodic layer
(Sec.~\ref{sec:setup}), \citet{TraorePerez2012} and \citet{Wu2013,Wu2015} follow
the convective state while decreasing $T$ and find that, for $C=M=10$, the
motion stops at $T_f=108.2$ with their finest grid. This corresponds to
$T_f/T_c=0.66$, against $T_{\rm NL}<109.73$ ($0.670\,T_c^h$) in our
computations. Both determinations locate the end of the same branch, by
continuation in their case and from the evolution of optimal perturbations in
ours. \citet{Wu2015} also show that $T_f$ is very sensitive to the numerical
diffusion of the charge equation. As in their simulations, the convective state
keeps $U_{\max}$ above the drift-velocity scale down to the end of the branch. The nonlinear
stability analysis of \citet{AttenLacroix1979}, which assumes $M\to\infty$ and a
modal velocity field, gives the higher values $125$ and $116$ for one and two
modes. The difference is due to the self-similar velocity field assumed in that
analysis \citep{Wu2015}.

\begin{figure*}
  \centering
  \includegraphics[width=\linewidth]{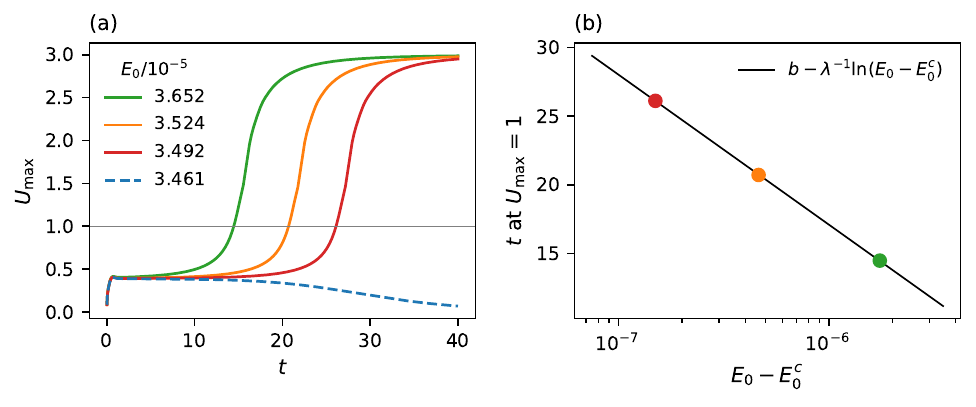}
  \caption{Approach to the minimal seed at $T=159.73$ ($\Delta T=4.1$).
    (a) Maximum velocity of the optimal trajectories for the largest
    non-transitioning amplitude of the bracket (dashed) and the three
    transitioning amplitudes (solid) used in (b).
    (b) Time at which $U_{\max}=1$ versus $E_0-E_0^c$, with the law of
    Eq.~\eqref{eq:edge}, $\lambda=0.211$. The colors match those of (a).}
  \label{fig:edge}
\end{figure*}

\paragraph{Edge state.}
For values of $T$ near the threshold ($\Delta T = 4.1$ in
Table~\ref{tab:minseed}), the bracket was narrowed to $0.9\%$ by integrating to
$t=40$, which reveals how trajectories behave close to the minimal seed
[Fig.~\ref{fig:edge}(a)]. All perturbations first relax within $t\approx1$ to a
common state of intermediate amplitude, $U_{\max}\approx0.4$, and stay near it.
Those above the critical amplitude eventually leave it towards the convective
state, and those below it decay. The time at which the transitioning
trajectories reach $U_{\max}=1$ diverges logarithmically as the amplitude
approaches the critical one [Fig.~\ref{fig:edge}(b)],
\begin{equation}
  \begin{aligned}
  &t_{\rm cross}\simeq b-\frac{1}{\lambda}\ln\!\left(E_0-E_0^c\right),\\
  &\lambda=0.211,\quad E_0^c=3.477\times10^{-5},
  \label{eq:edge}
  \end{aligned}
\end{equation}
where $\lambda$ is the growth rate measured independently in
Sec.~\ref{sec:modal}. With this value the three trajectories within $5\%$
of $E_0^c$ are fitted with residuals of $0.02$ time units, and a free
three-parameter fit gives $\lambda=0.214$.

This behavior identifies an edge state \citep{Skufca2006,Schneider2007}, a
saddle with a single unstable direction whose stable manifold separates the
basins of attraction of the hydrostatic and convective states. The minimal seed
is the point of smallest amplitude on this manifold [Fig.~\ref{fig:sketch}(b)].
A perturbation of amplitude $E_0$ starts at a distance $\delta\propto E_0-E_0^c$
from the manifold. Its stable components decay quickly, so it is carried along
the manifold to the neighborhood of the edge state, which produces the plateau.
There, its component along the unstable direction grows as $\delta\,e^{\lambda
t}$, so it leaves after a time $\lambda^{-1}\ln\left(\delta^{-1}\right)$
\citep[e.g.][chap.~6]{Strogatz2015}. This is the logarithm in Eq.~\eqref{eq:edge},
whose slope therefore measures the unstable eigenvalue. The time spent reaching
the plateau does not depend on $\delta$ and is absorbed into $b$.

\begin{figure*}
  \centering
  \includegraphics[width=\linewidth]{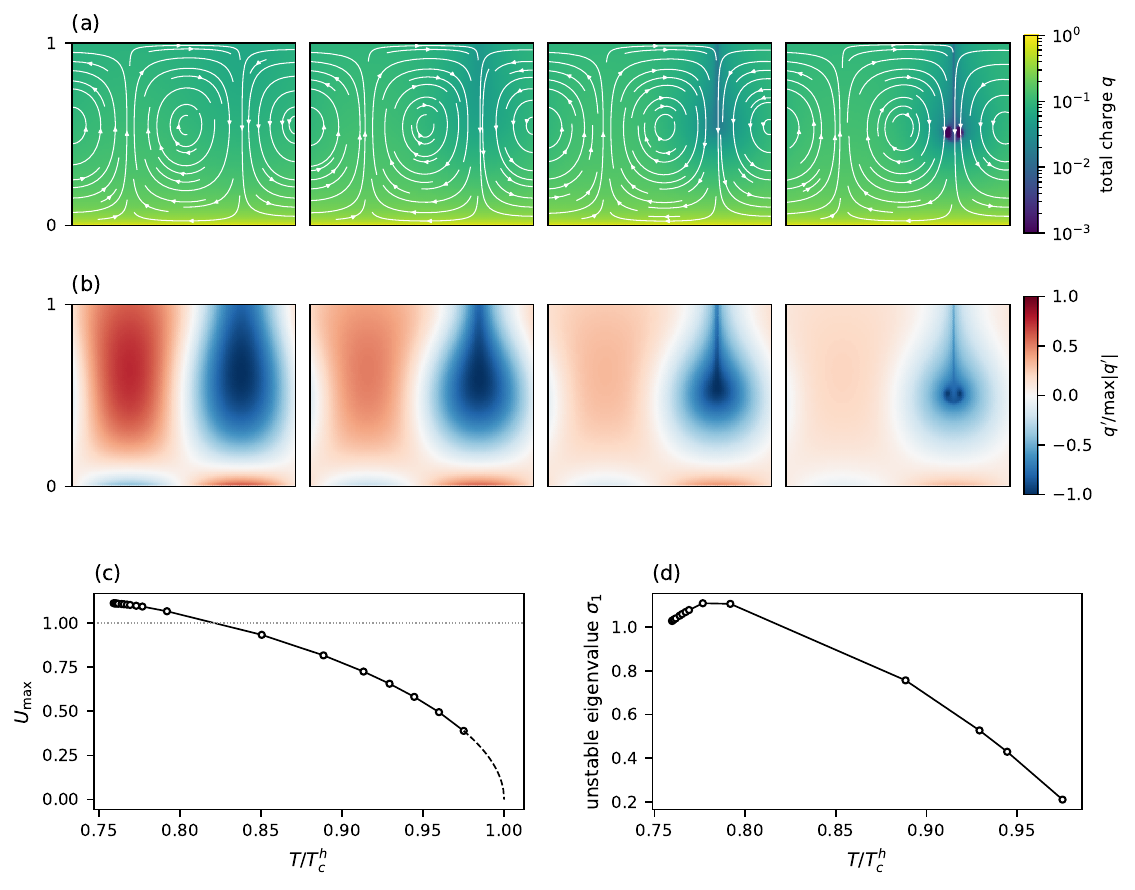}
  \caption{Computed edge states (production mesh). (a) Total charge
    $q_b+q'$ (logarithmic scale) and streamlines. (b) Charge perturbation
    $q'$, normalized by its maximum. From left to right, $T/T_c^h=0.975$,
    $0.888$, $0.792$ and $0.760$, with $U_{\max}=0.39$, $0.82$, $1.07$ and
    $1.11$ and $\max|q'|=0.016$, $0.044$, $0.081$ and $0.12$.
    (c) Maximum velocity along the branch, with the law
    $U_{\max}\propto(T_c^h-T)^{1/2}$ near threshold (dashed). The dotted
    line is the drift-velocity scale, $U_{\max}=1$. (d) Unstable eigenvalue of the edge
    state.}
  \label{fig:edge_states}
\end{figure*}

The edge state can be computed directly as a steady solution of the discrete
equations. Starting from the plateau of the bracketing trajectories at
$T=159.73$, Newton's method converges to a steady state, with $U_{\max}=0.389$
and $K=0.0268$, within $0.4\%$ of the plateau values. Its linearization has a
single unstable eigenvalue, $\sigma_1=0.2108$, which coincides with the escape
rate $\lambda=0.211$ measured from the trajectories, a zero eigenvalue
associated with horizontal translations, and a stable remainder
($\mathrm{Re}\,\sigma\le-2.9$). The edge state was followed in $T$ by
pseudo-arclength continuation, restricted to mirror-symmetric fields, down to
$T/T_c^h=0.76$ (Fig.~\ref{fig:edge_states}). Near threshold $U_{\max}^2\propto
T_c^h-T$, and extrapolating it to zero gives $T=163.78$, against the linear
value computed in Sec.~\ref{sec:linear}, $T_c^h=163.82$. The edge states form
the unstable branch of the subcritical pitchfork that leaves the hydrostatic
state at the linear threshold. Figure~\ref{fig:edge_states}(b) shows the evolution
of the edge states as $T$ moves away from the linear threshold. Close to it, the
edge state has the structure of the critical mode. Further from threshold the
rolls strengthen and the charge perturbation concentrates into a deficit, the
same sign asymmetry as the finite-amplitude optimal perturbations. The unstable
eigenvalue grows to $\sigma_1\simeq1.1$ near $T/T_c^h=0.78$ and then decreases,
as it must vanish at the end of the unstable branch when it approaches the
saddle node. When $U_{\max}$ exceeds the drift-velocity scale, near
$T/T_c^h=0.82$, a charge void forms inside each roll, where the charge, carried by
$\bu+\bE$, follows closed orbits and decays to zero. The void is displaced from
the center of the roll towards the down-flow, around the point where
$\bu+\bE=\bm0$. The spectrum then acquires a cluster
of weakly damped modes associated with the void. The continuation was
stopped at $T/T_c^h=0.76$ because Newton's method no longer converges reliably
with the stabilized discretization, which produces small negative charge
undershoots inside the void.

\begin{figure}
  \centering
  \includegraphics[width=\columnwidth]{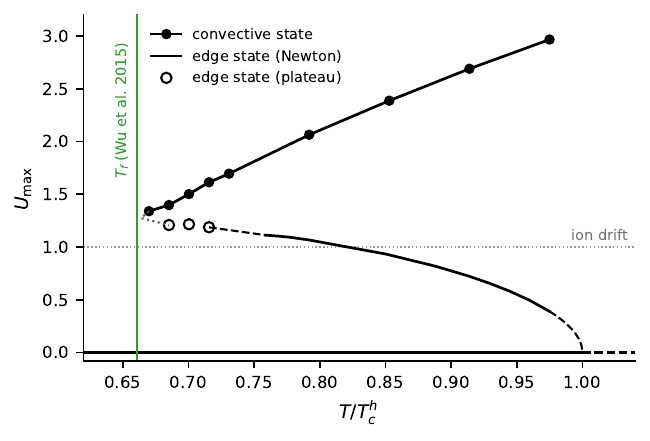}
  \caption{Computed bifurcation diagram on the production mesh, showing the maximum
  velocity of the convective state (filled symbols) and of the edge state,
  computed by Newton's method and pseudo-arclength continuation down to
  $T/T_c^h=0.76$ (solid line) and estimated from the plateau of the
  trajectories that bracket the critical amplitude within $4\%$ (open
  symbols). Dashed: pitchfork law near $T_c^h$ and interpolation between
  computed and estimated edge states. Dotted: schematic saddle-node.}
  \label{fig:branches}
\end{figure}

\paragraph{Bifurcation diagram.}
Figure~\ref{fig:branches} collects the convective and edge states. Below
$T/T_c^h=0.76$, where the continuation stops, the edge state can still be
estimated from the trajectories near the critical amplitude, wherever the
bisection bracket is narrow enough (at most $4\%$) for them to linger near
it. Its maximum velocity, read on the plateau of the smallest
transitioning trajectory, is $U_{\max}=1.19$--$1.22$ between
$T/T_c^h=0.716$ and $0.685$, continuing the computed branch, which reaches
$U_{\max}=1.11$ at $T/T_c^h=0.760$. Over the same range the convective state
weakens from $U_{\max}=1.61$ to $1.40$, and the two branches approach each other
and merge at the saddle node $T_{\rm NL}$, close to the finite-amplitude
criterion $T_f$ of \citet{Wu2015}. Both the convective state and the edge
state have $U_{\max}$ above the drift-velocity scale near the end of the branch.

\subsection{Modal anatomy of the optimal trajectories}
\label{sec:modal}

The trajectories can be decomposed on the eigenmodes of the linearized
dynamics at the same $T$. The discrete perturbation problem is the
differential--algebraic system $\mathsf M\dot{\bz}'+\mathcal S'(\bz')=0$
(Appendix~\ref{app:weak}), where $\mathsf M$ contains the time-derivative
terms of velocity and charge (including the SUPG term) and is singular in
$p'$ and $\phi'$. With $\mathsf J$ the Jacobian of $\mathcal S'$ at
$\bz'=0$, the temporal modes and the adjoint modes satisfy
\begin{equation}
  -\mathsf J\hat{\bz}_j=\sigma_j\mathsf M\hat{\bz}_j,\qquad
  -\mathsf J^{\mathsf T}\hat{\bz}^\dagger_j=\sigma_j\mathsf M^{\mathsf T}\hat{\bz}^\dagger_j ,
  \label{eq:temporal_modes}
\end{equation}
and are biorthogonal, $\hat{\bz}_i^{\dagger\mathsf T}\mathsf M\hat{\bz}_j=0$ for
$\sigma_i\neq\sigma_j$. Because the operator is non-normal the direct modes
are far from orthogonal, so a state is expanded with the direct modes, but using
the orthogonality of the adjoint modes to compute the coefficients,
\begin{equation}
  \bz'(t)=\sum_ja_j(t)\hat{\bz}_j+\bm r(t),\qquad
  a_j=\frac{\hat{\bz}_j^{\dagger\mathsf T}\mathsf M\bz'}
           {\hat{\bz}_j^{\dagger\mathsf T}\mathsf M\hat{\bz}_j},
  \label{eq:biorth}
\end{equation}
which is exact for each retained mode whatever the content of the rest of the
spectrum and gives $a_j(t)=a_j(0)e^{\sigma_jt}$ in the linear regime
\citep{SchmidHenningson2001,Schmid2007}. The forty modes with the largest
$\mathrm{Re}\,\sigma$ are computed by shift-and-invert Arnoldi on
Eq.~\eqref{eq:temporal_modes} with $\mathsf J$ and $\mathsf M$ assembled from the
residual of the time stepper. In the periodic layer they come in degenerate
pairs (translations of each other), and the kinetic energy $K_g$ of the component
of each pair is independent of the horizontal phase. At $T=159.73$ the least
stable pair is the $n=1$ roll, $\sigma_1=-0.1166$, consistent with the
slope $\mathrm d\sigma_{\rm NM}/\mathrm dT$ of Sec.~\ref{sec:threshold}. The
next $n=1$ pairs are oscillatory and strongly damped,
$\sigma=-5.42\pm9.08\mathrm i$ and $-6.84\pm14.56\mathrm i$.

\begin{figure}
  \centering
  \includegraphics[width=\linewidth]{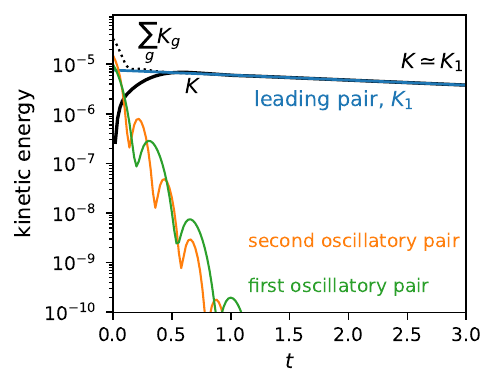}
  \caption{Linear optimum at $T=159.73$ ($E_0=10^{-8}$, $\tau=1$).
    Decomposition on the temporal eigenmodes, showing the kinetic energy $K$,
    the energy $K_1$ of the component of the leading $n=1$ pair, energies of the
    components of the first and second oscillatory $n=1$ pairs
    ($\sigma=-5.42\pm9.08\,\mathrm i$ and $-6.84\pm14.56\,\mathrm i$), and
    $\sum_gK_g$, which exceeds $K$ when the modal components cancel.}
  \label{fig:modal_linear}
\end{figure}

\paragraph{Linear optimum.} Figure~\ref{fig:modal_linear} shows the evolution in
time of the kinetic energy for a direct simulation using the linear optimum as
the initial condition ($E_0=10^{-8}$, $\tau=1$). At $t=0$ the velocity is zero,
yet the leading pair already carries a kinetic energy $K_1=7.7\times10^{-6}$.
This kinetic energy is cancelled by the two oscillatory $n=1$ pairs, whose
components carry $1.0\times10^{-5}$ and $1.5\times10^{-5}$.

Each oscillatory pair, $\sigma=-\gamma\pm\mathrm i\omega$, contributes to the
velocity a term proportional to $e^{-\gamma t}\cos(\omega t+\varphi)$. Its
kinetic energy oscillates at frequency $2\omega$ inside an envelope that decays
as $e^{-2\gamma t}$, which produces the lobes in Fig.~\ref{fig:modal_linear}.
The damping times, $1/\gamma=0.18$ and $0.15$, are shorter than the periods,
$2\pi/\omega=0.69$ and $0.43$, so the oscillatory components decay in less than
one period. As they decay, the cancellation disappears and the leading mode is
unmasked. The kinetic energy exceeds that of the leading pair by at most $4\%$,
at $t\simeq0.65$, and the difference $K-K_1$ changes sign every half period
$\pi/\omega$ of the first oscillatory pair. Because the modes are not
orthogonal, $K$ is not the sum of the modal energies. It also contains cross
terms, of order $2\sqrt{K_1K_{\rm osc}}$, which are much larger than the energy
$K_{\rm osc}$ of the oscillatory components themselves. The truncation of the
expansion plays no role here, since for $t>0.55$ the residual $\bm r$ carries
less than $10^{-4}$ of the kinetic energy. The oscillatory modes are
transit-time modes of the ion transport, and they are discussed in detail in
Sec.~\ref{sec:transit}.

The kinetic energy rises to its maximum
$6.9\times10^{-6}$ at $t=0.58$, after which $K$ approaches $K_1$ and decays at
the rate $2\sigma_1$. The optimum maximizes the leading-mode content per unit charge
energy ($K_1/E_0=765$) while making the initial velocity vanish by superposing
non-orthogonal fast modes. This is the mechanism of non-modal growth
\citep{Trefethen1993}.

\begin{figure*}
  \centering
  \includegraphics[width=\linewidth]{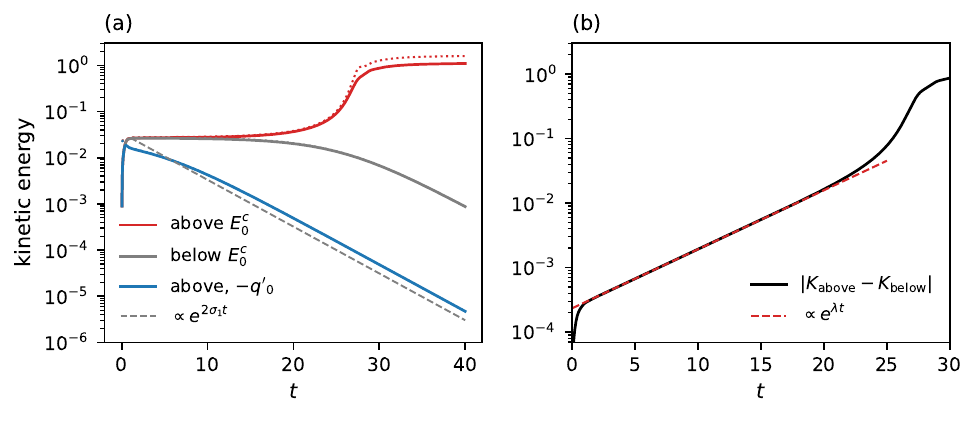}
  \caption{Optimal trajectories at the critical amplitude, $T=159.73$. (a)
    Optimal perturbations just above ($E_0=3.492\times10^{-5}$) and just below
    ($E_0=3.461\times10^{-5}$) the critical amplitude, and sign reversal of the
    first, showing the kinetic energy $K$ (solid) and the energy of the leading pair (dotted).
    Dashed: linear decay from $t=1$. (b) Separation of the trajectories above
    and below the critical amplitude, with an exponential fit
    $\propto e^{\lambda t}$, $\lambda=0.211$.}
  \label{fig:modal_seed}
\end{figure*}

\paragraph{Minimal seed.} Figure~\ref{fig:modal_seed}(a) shows the optimal
perturbations just above and just below the critical amplitude
($E_0=3.492\times10^{-5}$ and $3.461\times10^{-5}$, $0.9\%$ apart) and the
sign reversal of the first. Up to $t\approx0.6$ the evolution repeats that
of the linear optimum, with $K$ rising to $0.027$ and $U_{\max}$ to $0.39$.
Then, between $t\approx1$ and $15$, both trajectories stay on a plateau
where $K$ hardly changes, with $K(10)/K(1)=1.06$ above and $0.99$ below the
critical amplitude, against $0.12$ for linear decay. The leading $n=1$ pair
accounts for the plateau energy to within $3\%$ ($K_1/K=1.03$). The rest of
the expansion consists of $n=2$ harmonics forced quadratically and a residual
outside the retained modes with $6\%$ of the kinetic energy.
The edge state is therefore, to within a few percent, a finite-amplitude
version of the least stable linear mode, in which the nonlinear terms
balance its linear decay. The two trajectories separate exponentially,
\begin{equation}
  |K_{\rm above}-K_{\rm below}|\propto e^{\lambda t},\qquad\lambda=0.211 ,
\end{equation}
with $\lambda$ between $0.210$ and $0.213$ on every window between $t=2$ and
$18$ [Fig.~\ref{fig:modal_seed}(b)]. This $\lambda$ is the growth rate of the single
unstable direction of the edge state, the value used in Eq.~\eqref{eq:edge}.
Above the critical amplitude the trajectory leaves the space spanned by the
linear modes (on the convective state the residual $\bm r$ carries seven
times the kinetic energy of the flow, so the expansion loses its meaning),
whereas below it returns to the linear regime and decays at $2\sigma_1$.

\paragraph{Sign reversal.} The linear part of the expansion is identical for
$\pm q'_0$, and the leading-pair energy at $t=0$ is $K_1=2.38\times10^{-2}$ in
both cases. The difference is produced by the nonlinear terms during the
initial transient. At $t=1$ the optimum has raised $K_1$ to
$2.71\times10^{-2}$, enough to reach the edge state, whereas its sign
reversal has lowered it to $1.64\times10^{-2}$ and decays without
plateau, eventually at the linear rate. The charge-deficit asymmetry of
Sec.~\ref{sec:mechanism} thus acts within the first time unit, by setting
how much energy reaches the mode that leads to the edge state. The
minimal seed puts about $11\%$ less energy per unit $E_0$ into the leading
mode than the linear optimum ($K_1/E_0=682$ against $765$), trading linear
efficiency for the sign-asymmetric structure that gains energy
nonlinearly.

\subsection{Implications for the onset of electroconvection}
\label{sec:onset}

These results bear directly on the discrepancy between the linear threshold
and the experimental onset of electroconvection. Between $T_c$ and
$T_{\rm NL}\approx0.67\,T_c$ the hydrostatic state is destabilized by charge
perturbations whose root-mean-square value is a few percent of the injected
charge, and the most efficient such perturbations are concentrated charge
deficits. Such fluctuations are plausible in experiments, for example from
inhomogeneous injection through ion-exchange membranes \citep{Feng2021}. Finite-amplitude triggering can therefore lower the
onset of convection from $T_c$ to about two thirds of it, but not below the
end of the convective branch. The experimental onset $T\approx100$
($T/T_c\approx0.6$) reported by \citet{Lacroix1975} lies somewhat lower.
Charge diffusion, which lowers the stability thresholds
\citep{PerezCastellanos1989}, and three-dimensional or finite-size effects in
the experiments are possible reasons for the remaining difference.

\begin{figure*}
  \centering
  \includegraphics[width=\linewidth]{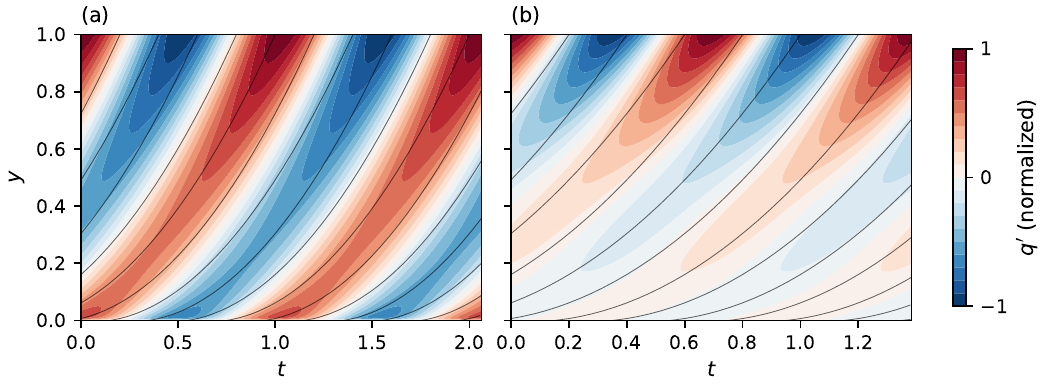}
  \caption{Charge perturbation of the least damped oscillatory modes in the
    $(t,y)$ plane, $C=10$, with the decay factor removed. (a) Horizontally
    uniform mode ($n=0$, no flow), exact solution, $\sigma=-3.01\pm6.09\,\mathrm i$.
    (b) First oscillatory $n=1$ mode of the complete problem ($M=10$,
    $T=159.73$, $k=2\pi/L$, spectral code of Sec.~\ref{sec:transit}), $\sigma=-5.36\pm9.08\,\mathrm i$, at the
    horizontal position of maximum amplitude. Thin lines: paths of the ions of
    the hydrostatic state, $t-\int_0^y\dd y'/E_b=\text{const}$.}
  \label{fig:transit}
\end{figure*}

\section{Oscillatory transit-time modes}
\label{sec:transit}

The decomposition of Sec.~\ref{sec:modal} shows that, after the leading roll
mode, the least damped $n=1$ modes are oscillatory pairs, and that non-modal
growth relies on them. Their existence bears on a question left open by
\citet{AttenMoreau1972}. For weak injection they proved the exchange of
stabilities. Oscillatory perturbations can exist, but always with a negative
growth rate. For arbitrary injection they could not extend the proof, searched
numerically for $C=5$ for neutral oscillatory solutions $\sigma=\mathrm i\beta$
($0\le T\le2000$, $2\le k\le12$, $0\le\beta\le50$), and, finding none,
concluded that the instability sets in as a steady cellular motion. In this
section we identify the oscillatory modes, explain their origin and show that
they remain damped for $C=10$.

\subsection{Spectral computation}

The temporal spectrum of the linearized problem is computed for a single
horizontal Fourier mode,
$(\bu',p',\phi',q')=(\hat{\bu},\hat p,\hat\phi,\hat q)(y)\,
e^{\mathrm ikx+\sigma t}$, which reduces it to a one-dimensional generalized
eigenvalue problem for the vertical velocity, the charge density and the
electric potential. The vertical structure is discretized by Chebyshev
collocation with $N=60$, and the complete spectrum is obtained with a dense QZ
solver. This code gives $T_c=164.06$ and
$k_c=5.113$ ($\lambda_c=1.229$), against $T_c\simeq164.1$ and $k_c=5.113$ of
\citet{AttenMoreau1972} and $T_c=164.09$ of \citet{Perez2014}. At $T=159.73$
and $k=2\pi/L$ it gives $\sigma_1=-0.124$ and the oscillatory pairs
$-5.36\pm9.08\mathrm i$ and $-6.72\pm14.49\mathrm i$. The frequencies agree
with those of the two-dimensional discrete operator of Sec.~\ref{sec:modal} to
within $0.5\%$. The difference in $\sigma_1$ reflects the threshold of the
production mesh, $T_c^h=163.82$, against $164.06$ for the converged problem.
Increasing $N$ to $90$ changes these eigenvalues by less than $10^{-3}$.
Because the problem separates in $x$, a complete spectrum costs about
$10^{-2}$\,s, several orders of magnitude less than shift-and-invert Arnoldi on
the two-dimensional operator, which makes it possible to explore the
$(k,T,M)$ parameter space systematically.

\subsection{Physical origin of the oscillatory modes}

The physical origin of the oscillatory modes is the transport of charge by the
electric field. Dividing the momentum equation, Eq.~\eqref{eq:ns}, by $M^2$ we
can see that the inertia disappears as $M\to\infty$. The oscillatory pairs
persist in this limit, with frequencies converging to $9.59$ and $15.12$
(against $9.08$ and $14.49$ for $M=10$). Horizontally uniform modes ($n=0$),
which carry no flow, are oscillatory as well, with frequencies independent of
$M$. For them the linearized problem can be solved exactly along the ion paths
(Appendix~\ref{app:transit}), and their growth rates are the roots of
\begin{equation}
  P(\sigma)=C^2e^{-\sigma t_{\rm tr}},
  \label{eq:transit_main}
\end{equation}
where $P$ is a cubic polynomial and $t_{\rm tr}=\int_0^1\dd y/E_b=1.248$ is the
transit time of the ions across the layer. The roots form an infinite sequence
of damped modes with frequencies spaced by about $2\pi/t_{\rm tr}$. The first is
$-3.014\pm6.092\mathrm i$. The spectral code reproduces them with a relative
error below $3\times10^{-5}$ (Table~\ref{tab:transit}). Each mode is a charge
wave carried from the injector to the collector at the drift velocity, and the
delay of one transit sets its frequency [Fig.~\ref{fig:transit}(a)]. The field
perturbation created at the injector is carried along the ion paths, and when it
leaves the layer at the collector, one transit later, the fixed applied voltage
requires a compensating change of the current, which starts a new perturbation
at the injector. The charge waves of the oscillatory $n=1$ modes travel along
the same ion paths [Fig.~\ref{fig:transit}(b)].

\begin{figure*}
  \centering
  \includegraphics[width=\linewidth]{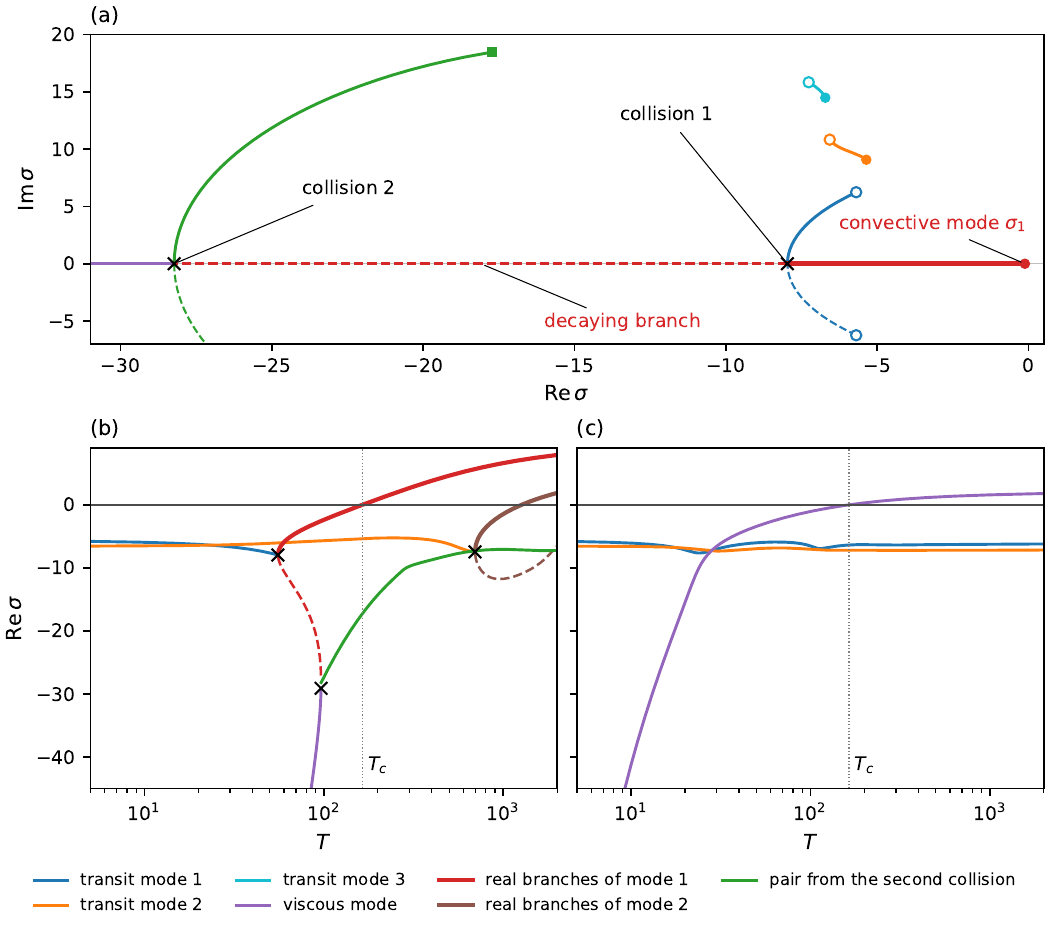}
  \caption{Continuation in $T$ of the first transit-time modes and of the
    slowest viscous mode, $C=10$, $k=2\pi/L$ (Chebyshev, $N=60$). (a)
    Eigenvalues in the complex plane for $M=10$, as $T$ increases from $1$
    (open circles) to $159.73$ (filled symbols). The second and third transit
    modes become the oscillatory pairs $-5.36\pm9.08\,\mathrm i$ and
    $-6.72\pm14.49\,\mathrm i$, whose conjugates are not shown. Dashed:
    conjugate of the first mode, and the decaying real branch. After the first
    collision ($T=55.6$) the two real branches are the convective mode (thick
    line) and a decaying branch, which at the second collision ($T=96.6$)
    merges with the slowest viscous mode into a strongly damped pair (square at
    $T=159.73$). (b), (c) Real part of the eigenvalues as a function of $T$ for
    $M=10$ and $M=3$. $\times$: points where two eigenvalues meet. After a pair
    reaches the real axis its two real branches are drawn in a new color, solid
    for the one that rises and dashed for the one that moves down. Dotted:
    $T_c=164.06$.}
  \label{fig:continuation_T}
\end{figure*}

\subsection{From transit modes to the convective mode}
\label{sec:transit_coupling}

Below $T_c$ the base state is at rest, and the velocity appears only as a
component of the perturbation modes. In a mode, the velocity perturbation that
accompanies the charge perturbation follows from the balance between the
viscous and Coulomb forces in Eq.~\eqref{eq:ns},
$(M^2/T)\nabla^2\bu'\sim CM^2q'\bE$, and is of order $TC|q'|$ in units of the
drift velocity. For $T$ well below $T_c$ this velocity component is small and
hardly modifies the charge transport, and the modes are close to the transit
modes of the $n=1$ electrical problem, in which the velocity perturbation is
suppressed. For
$k=2\pi/L$ these are $-5.67\pm6.27\mathrm i$, $-6.58\pm10.85\mathrm i$ and
$-7.27\pm15.85\mathrm i$. They are close in frequency to the $n=0$ modes, but
more damped, because the field perturbation is screened over a distance $1/k$.
At $M=10$ and $T=10$ the first two pairs, $-5.94\pm5.92\mathrm i$ and
$-6.52\pm10.65\mathrm i$, differ from them by $5\%$ and $2\%$, and the deviation
grows linearly with $T$. As $T$ increases at fixed $M$, which is the path
followed in an experiment when the voltage is raised, the coupling between the
charge and velocity perturbations becomes important.

Figure~\ref{fig:continuation_T}(a) follows the modes from $T=1$ to $T=159.73$ for
$M=10$ and $k=2\pi/L$. The second and third modes (orange and cyan in the
figure) become the oscillatory pairs $-5.36\pm9.08\mathrm i$ and
$-6.72\pm14.49\mathrm i$, with frequencies lowered by $16\%$ and $9\%$ with
respect to the electrical problem. The first mode becomes more damped and its
frequency decreases. At $T=55.6$ its imaginary part vanishes, and the two
complex-conjugate eigenvalues of the pair meet on the real axis
($\sigma\simeq-7.97$) and split into two real modes. One of them becomes the
convective mode, $\sigma_1=-0.124$ at $T=159.73$. The other moves to the left
and, at $T=96.6$ ($\sigma\simeq-28.2$), meets the least damped viscous mode of
the flow. Without coupling, this mode is the slowest viscous decay of a flow
perturbation between rigid walls. Its growth rate is $\sigma=-(M^2/T)\mu_1$,
where $\mu_1$ is the smallest eigenvalue of
$(\mathrm D^2-k^2)^2\psi=-\mu(\mathrm D^2-k^2)\psi$ with $\psi=\mathrm D\psi=0$
at $y=0,1$, and $\mathrm D=\dd/\dd y$. It depends only on $k$, and
$\mu_1=47.55$ for $k=2\pi/L$. The viscous mode is thus strongly damped for small
$T$ and approaches the transit modes as $T$ increases. After the second
collision the two form a strongly damped oscillatory pair,
$-17.73\pm18.46\mathrm i$ at $T=159.73$.

Figure~\ref{fig:continuation_T}(b) shows the real parts of the same modes as
functions of $T$, up to $T=2000$. The collision of the first pair
and that of the decaying branch with the viscous mode are the left and middle
crosses. The second pair behaves in the same way at higher $T$. Its two
eigenvalues meet on the real axis at $T=698$ ($\sigma\simeq-7.46$, right
cross), and one of the resulting real branches becomes unstable at
$T=1267=7.7\,T_c$. The collision of the first pair moves up to $T=61.8$ for
$M=30$ and to $62.6$ in the inertia-free limit, and down to $T=40.2$
($\sigma\simeq-10.5$) for $M=5$. For smaller $M$ the sequence is different. For
$M=3$ [Fig.~\ref{fig:continuation_T}(c)] the first transit pair remains oscillatory
up to $T_c$ ($-6.37\pm7.28\mathrm i$ there), and the mode that becomes unstable
is the slowest viscous mode, which the Coulomb coupling makes less and less
damped as $T$ increases ($\sigma=-15.5$ at $T=20$, against $-(M^2/T)\mu_1=-21.4$
without coupling) until it crosses zero. Since $M$ drops out of the marginal
problem, the marginal mode at $T_c$ is the same in all cases. What depends on
inertia is the branch of the stable spectrum from which it is reached.

The change of branches occurs at $M^*\simeq 4.8$. This can be understood by
comparing the damping rates of the two branches. The slowest viscous mode decays
at the rate $(M^2/T)\mu_1$, where $M^2/T=\nu/(KV_0)$ is the ratio of the drift
time to the viscous time, the inverse of the parameter $R=KV_0/\nu$ of
\citet{AttenMoreau1972}. At $T_c$ this rate is $2.6$ for $M=3$, $6.5$ for
$M=4.75$, $7.2$ for $M=5$ and $29$ for $M=10$. The first transit mode decays at
a rate of about $6$--$8$, set by the transit of the ions. The Coulomb feedback
destabilizes the less damped of the two. For small $M$ the flow is the slow
variable and the charge adapts to it, so the instability grows out of the
viscous mode. For large $M$ the flow adapts quickly and the slow variable is the
charge transport, so the instability grows out of a transit mode. The same
happens in Rayleigh--B\'enard convection, where the Prandtl number decides
whether the critical mode continues the slowest viscous or the slowest thermal
mode, although there both modes are real and need not collide
\citep{Chandrasekhar1961}.

\subsection{Exchange of stabilities}

\begin{figure}
  \centering
  \includegraphics[width=\columnwidth]{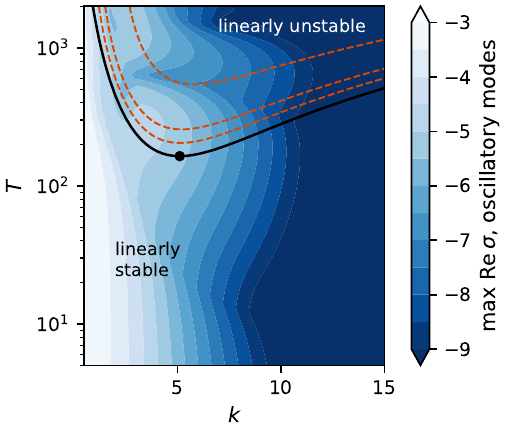}
  \caption{Least damped oscillatory eigenvalue in the $(k,T)$ plane, $C=10$,
    $M=10$ (Chebyshev, $N=60$). Colors: its real part, negative everywhere. The
    real eigenvalue of the steady mode is not included. Black line: neutral curve
    of the steady mode, above which the base state is linearly unstable through
    this real eigenvalue. Dashed: its growth rate, $\sigma=1$, $2$ and $5$. Dot:
    $(k_c,T_c)$.}
  \label{fig:scan}
\end{figure}

The continuation of Sec.~\ref{sec:transit_coupling} shows that, at $k=2\pi/L$, the
mode that becomes unstable is real for all the values of $M$ considered. To
check whether any oscillatory mode becomes unstable, the spectrum was computed
on a grid of $59$ wavenumbers $0.5\le k\le15$ and $90$ values $5\le T\le2000$,
for $M=1$, $3$, $10$, $30$, $100$ and in the inertia-free limit, about
$3\times10^4$ spectra in a few minutes. No oscillatory mode becomes unstable.
Over the whole grid and for every $M$, the least damped one has
$\mathrm{Re}\,\sigma\le-3.0$. It is always found at small wavenumbers,
$k\le0.75$, where the modes approach the horizontally uniform transit modes, and
near $k_c$ the oscillatory modes stay between $\mathrm{Re}\,\sigma=-7.2$ and
$-4.5$ (Fig.~\ref{fig:scan}). For $C=10$ and the range explored by
\citeauthor{AttenMoreau1972}, the instability therefore sets in through a real
eigenvalue, and the continuation of Sec.~\ref{sec:transit_coupling} explains
why. Each steady mode is either a transit-time pair made real by the flow
feedback ($M>M^*$) or the slowest viscous mode destabilized by it ($M<M^*$),
while the remaining pairs stay strongly damped. This is a numerical result for
one injection strength, not a proof of the exchange of stabilities.

\FloatBarrier
\section{Conclusions}
\label{sec:conclusions}

We have formulated and validated a nonlinear non-modal optimization framework
for electroconvection driven by unipolar injection between two infinite
parallel plates, modeled by a periodic domain of one critical wavelength. We have used a monolithic
finite-element discretization in perturbation variables and an automatically
derived discrete adjoint. Writing the discrete problem as the difference between the residuals
of the perturbed and base states makes the zero perturbation an exact discrete
solution, and Taylor tests confirm that the gradient is exact in the linear and
nonlinear regimes. The initial perturbation on the injected charge is
constrained by its energy, and the objective functional is the kinetic energy at
a given horizon. 

In the small-amplitude, long-horizon limit the asymptotic growth rate of the
optimal response reproduces the linear threshold. The asymptotic growth rate
vanishes at $T_c^{\rm NM}=163.827$, which coincides with
the discrete modal threshold of the same discretization to within $0.003\%$.
The modal threshold converges at second order to $164.06$, within $0.02\%$
of the published values \citep{AttenMoreau1972,Perez2014}.

Below the linear threshold, nonlinear effects on the optimal perturbation become
noticeable for an amplitude of the initial perturbation of $E_0\gtrsim10^{-6}$
and strong for $E_0\sim10^{-4}$, where the optimal gain increases by a factor of
up to four for the longest horizons and the optimal initial charge differs
structurally from the linear one. The finite-amplitude optimum remains
mirror-symmetric but becomes strongly asymmetric in sign, concentrating into a
charge deficit with a horizontally uniform component and a second harmonic. This
follows from the amplitude-dependent Coulomb relaxation rate $C(2q_b+q')$. 

The minimal seed grows monotonically with the distance to threshold, from
$E_0^c=3.5\times10^{-5}$ at $T/T_c=0.975$ to $(6.5$--$7.5)\times10^{-4}$ at
$T/T_c=0.73$ and $1.1\times10^{-3}$ at $T/T_c=0.69$, i.e.,
root-mean-square charge perturbations of $0.5$--$3\%$ of the injected
charge, converged in resolution to the precision of the bracketing. The
convective state reached weakens as $T$ decreases and the convective branch
ends near $T_{\rm NL}\approx0.67\,T_c$, where it meets the branch of edge
states. This end of the branch agrees with the finite-amplitude
criterion obtained by continuation of the convective state in direct
numerical simulations \citep{Wu2015}.

A decomposition of the trajectories on the direct and adjoint temporal
eigenmodes explains these dynamics. Linear transient growth unmasks the
least stable roll mode, initially cancelled by fast oscillatory modes. Near
the critical amplitude, trajectories linger on an edge state that is a
finite-amplitude version of that mode, with a single unstable direction of
growth rate $\lambda=0.21$. The time to transition diverges as
$\lambda^{-1}\ln(E_0-E_0^c)^{-1}$, and the sign of the perturbation decides,
within the first time unit, whether the edge state is reached.
Finite-amplitude triggering by small charge fluctuations thus lowers the
onset of electroconvection to about two thirds of the linear threshold,
accounting for most of the gap between the linear threshold and the
experimental onset.

The fast oscillatory modes involved in this growth are transit-time modes of
the ion drift. For horizontally uniform perturbations they are the roots of
an exact dispersion relation, $P(\sigma)=C^2e^{-\sigma t_{\rm tr}}$, with
frequencies spaced by about $2\pi/t_{\rm tr}$. As $T$ increases towards $T_c$,
the coupling between charge and velocity perturbations grows and the first of
them becomes the steady
convective mode for $M=10$, and the others remain oscillatory and strongly damped. A scan of the spectrum for
$C=10$, $T\le2000$ and $0.5\le k\le15$ finds no unstable oscillatory mode,
which supports, for strong injection, the exchange of stabilities proved by
Atten and Moreau for weak injection. Which branch of the stable spectrum leads
to the convective mode depends on the mobility parameter. For $M>M^*\simeq4.8$
it is the first transit mode, and for smaller $M$ the slowest viscous mode. The
change of branch takes place far below threshold, at $T\approx40$, and has no
observable effect near it. Liquids, with $M$ of order $10$ or larger, are in the first regime,
whereas gases, with $M\sim10^{-2}$, are deep in the second. Since the
non-modal growth found here relies on the transit modes, the transient
dynamics below threshold may differ in gases, where charge diffusion,
neglected here, also becomes more important.

A natural continuation of this work is to include charge diffusion, which is
known to lower the linear threshold. Three-dimensional effects may also
lower the onset of convection.

\section*{Supplementary Material}
See the supplementary material for the Python scripts used in this work,
with documentation, a license file, and instructions to reproduce the results.
The scripts include the Firedrake implementation of the direct--adjoint
optimization of the nonlinear electrohydrodynamic equations
(Secs.~\ref{sec:optimization}--\ref{sec:nonlinear} and
Appendices~\ref{app:adjoint}--\ref{app:weak}), the computation of edge
states by Newton's method and pseudo-arclength continuation, and the
Chebyshev spectral code for the transit-time modes and their continuation in
$T$ (Sec.~\ref{sec:transit} and Appendix~\ref{app:transit}). A short check
that reproduces values of Tables~\ref{tab:modal}, \ref{tab:spatial}
and~\ref{tab:transit} is described in the accompanying README file.

\begin{acknowledgments}
The authors acknowledge financial support from
MCIN/\allowbreak AEI/\allowbreak 10.13039/\allowbreak 501100011033/\allowbreak FEDER, UE (Grant No. PID2022-138890NB-I00). During
preparation of this manuscript, the authors used Claude (Anthropic,
Opus 5.5) to assist with writing scripts, analyzing data, and correcting
the manuscript's grammar. The authors reviewed the results and take full
responsibility for the final content. 

\end{acknowledgments}

\section*{Author Declarations}
\subsection*{Conflict of Interest}
The authors have no conflicts to disclose.

\subsection*{Author Contributions}
\textbf{Pedro A. V\'azquez:} Conceptualization (equal); Methodology (lead);
Software (lead); Formal analysis (lead); Investigation (lead); Visualization
(lead); Writing -- original draft (lead); Writing -- review \& editing (equal).
\textbf{A. T. P\'erez:} Conceptualization (equal); Formal analysis
(supporting); Funding acquisition (lead); Writing -- review \& editing (equal).

\section*{Data Availability Statement}
The data that support the findings of this study are available from the
corresponding author upon reasonable request.
The scripts used for the nonlinear non-modal optimization and for the
spectral computation of the transit-time modes are provided as supplementary
material.

\appendix
\begin{widetext}
\section{Derivation of the adjoint equations}
\label{app:adjoint}

This appendix derives the continuous adjoint equations associated with the
nonlinear optimization problem of Sec.~\ref{sec:optproblem}. The computational domain is
$\Omega$, with boundary $\partial\Omega=\Gamma_i\cup\Gamma_c\cup\Gamma_s$. The
nondimensional direct equations are Eqs.~\eqref{eq:ns}--\eqref{eq:poisson} with
the boundary conditions of Eq.~\eqref{eq:bc}. The control variable is the initial
charge perturbation constrained by Eq.~\eqref{eq:E0}, the objective functional is given by
Eq.~\eqref{eq:J}, and the augmented Lagrangian by Eq.~\eqref{eq:lagrangian}. The
adjoint equations, boundary conditions and terminal conditions are obtained
by requiring the first variation of $\mathcal{L}$ to vanish for arbitrary
admissible variations of the direct variables $\bu$, $p$, $q$ and $\phi$. The
derivation extends the procedure described by \citet{Kerswell2018} for the
Navier--Stokes equations. For concision we write
\begin{equation}
  \nu=M^2/T,\qquad \beta=CM^2 .
  \label{eq:nu_beta}
\end{equation}

The variation of the Lagrangian is
\begin{equation}
  \delta\mathcal{L}=\mathcal{L}(\bu+\delta\bu,p+\delta p,q+\delta q,\phi+\delta\phi)
  -\mathcal{L}(\bu,p,q,\phi).
  \label{eq:dL}
\end{equation}
Neglecting second-order terms,
\begin{align}
  \delta\mathcal{L}=&\int_\Omega\bu(\tau)\cdot\delta\bu(\tau)\dd\Omega
  -2\lambda\int_\Omega q'_0\,\delta q'_0\dd\Omega \nonumber\\
  &-\int_0^\tau\!\!\int_\Omega\ua\cdot\Big[\partial_t\delta\bu
    +(\delta\bu\cdot\nabla)\bu+(\bu\cdot\nabla)\delta\bu+\nabla\delta p\nonumber\\
  &\qquad-\nu\nabla^2\delta\bu+\beta\,\delta q\,\nabla\phi+\beta q\,\nabla\delta\phi
    \Big]\dd\Omega\dd t\nonumber\\
  &-\int_0^\tau\!\!\int_\Omega\pa\,\nabla\cdot\delta\bu\dd\Omega\dd t
  -\int_0^\tau\!\!\int_\Omega\qa\Big\{\partial_t\delta q
    +\nabla\cdot\Big[\delta q\,(\bu-\nabla\phi)\nonumber\\
  &\qquad+q\,(\delta\bu-\nabla\delta\phi)\Big]\Big\}\dd\Omega\dd t\nonumber\\
  &-\int_0^\tau\!\!\int_\Omega\phia\left(\nabla^2\delta\phi+C\,\delta q\right)\dd\Omega\dd t.
  \label{eq:dL_expanded}
\end{align}
The integrals are now manipulated by integration by parts, term by term.
Throughout, $\bn$ is the outward unit normal, $\partial_n=\bn\cdot\nabla$,
indices are summed when repeated, and $\big[(\nabla\bu)^{\mathsf T}\cdot\ua\big]_j
=u^{\dagger}_i\,\partial_ju_i$.

\subsection{Integration by parts}

\paragraph{Navier--Stokes terms.}
The time derivative gives
\begin{equation}
  -\int_0^\tau\!\!\int_\Omega\ua\cdot\partial_t\delta\bu\dd\Omega\dd t
  =-\int_\Omega\ua(\tau)\cdot\delta\bu(\tau)\dd\Omega
   +\int_\Omega\ua(0)\cdot\delta\bu(0)\dd\Omega
   +\int_0^\tau\!\!\int_\Omega\partial_t\ua\cdot\delta\bu\dd\Omega\dd t .
\end{equation}
The first convective term is already in the desired form,
\begin{equation}
  -\int_0^\tau\!\!\int_\Omega\ua\cdot(\delta\bu\cdot\nabla)\bu\dd\Omega\dd t
  =-\int_0^\tau\!\!\int_\Omega\delta\bu\cdot\big[(\nabla\bu)^{\mathsf T}\cdot\ua\big]
   \dd\Omega\dd t ,
\end{equation}
while the second one, using $\nabla\cdot\bu=0$, becomes
\begin{equation}
  -\int_0^\tau\!\!\int_\Omega\ua\cdot(\bu\cdot\nabla)\delta\bu\dd\Omega\dd t
  =-\int_0^\tau\!\!\oint_{\partial\Omega}(\bu\cdot\bn)(\ua\cdot\delta\bu)\dd S\dd t
   +\int_0^\tau\!\!\int_\Omega\delta\bu\cdot(\bu\cdot\nabla)\ua\dd\Omega\dd t .
\end{equation}
For the pressure and continuity terms,
\begin{align}
  -\int_0^\tau\!\!\int_\Omega\ua\cdot\nabla\delta p\dd\Omega\dd t
  &=-\int_0^\tau\!\!\oint_{\partial\Omega}(\ua\cdot\bn)\,\delta p\dd S\dd t
   +\int_0^\tau\!\!\int_\Omega\delta p\,\nabla\cdot\ua\dd\Omega\dd t ,\\
  -\int_0^\tau\!\!\int_\Omega\pa\,\nabla\cdot\delta\bu\dd\Omega\dd t
  &=-\int_0^\tau\!\!\oint_{\partial\Omega}\pa\,(\delta\bu\cdot\bn)\dd S\dd t
   +\int_0^\tau\!\!\int_\Omega\delta\bu\cdot\nabla\pa\dd\Omega\dd t ,
\end{align}
and Green's second identity applied to the viscous term gives
\begin{align}
  \nu\int_0^\tau\!\!\int_\Omega\ua\cdot\nabla^2\delta\bu\dd\Omega\dd t
  &=\nu\int_0^\tau\!\!\oint_{\partial\Omega}\left(\ua\cdot\partial_n\delta\bu
    -\delta\bu\cdot\partial_n\ua\right)\dd S\dd t\nonumber\\
   &\quad+\nu\int_0^\tau\!\!\int_\Omega\delta\bu\cdot\nabla^2\ua\dd\Omega\dd t .
\end{align}

\paragraph{Force terms in the momentum equation.}
The term
\begin{equation}
  -\beta\int_0^\tau\!\!\int_\Omega(\ua\cdot\nabla\phi)\,\delta q\dd\Omega\dd t
\end{equation}
is already in the desired form. The other term is integrated by parts,
\begin{equation}
  -\beta\int_0^\tau\!\!\int_\Omega(q\ua)\cdot\nabla\delta\phi\dd\Omega\dd t
  =-\beta\int_0^\tau\!\!\oint_{\partial\Omega}q\,(\ua\cdot\bn)\,\delta\phi\dd S\dd t
  +\beta\int_0^\tau\!\!\int_\Omega\nabla\cdot(q\ua)\,\delta\phi\dd\Omega\dd t ,
\end{equation}
where the divergence theorem has been used for the surface integral.

\paragraph{Charge equation.}
For the time derivative,
\begin{equation}
  -\int_0^\tau\!\!\int_\Omega\qa\,\partial_t\delta q\dd\Omega\dd t
  =-\int_\Omega\qa(\tau)\,\delta q(\tau)\dd\Omega
   +\int_\Omega\qa(0)\,\delta q(0)\dd\Omega
   +\int_0^\tau\!\!\int_\Omega\partial_t\qa\,\delta q\dd\Omega\dd t ,
\end{equation}
where $\qa(\tau)$ is short for $\qa(\bm{x},\tau)$, etc. For the first
advection term,
\begin{equation}
\begin{split}
  -\int_0^\tau\!\!\int_\Omega\qa\,\nabla\cdot\left[\delta q\,(\bu-\nabla\phi)\right]\dd\Omega\dd t
  =&-\int_0^\tau\!\!\oint_{\partial\Omega}\qa\,[(\bu-\nabla\phi)\cdot\bn]\,\delta q\dd S\dd t\\
  &+\int_0^\tau\!\!\int_\Omega(\bu-\nabla\phi)\cdot\nabla\qa\,\delta q\dd\Omega\dd t .
\end{split}
\end{equation}
The second advection term contains one contribution in $\delta\bu$,
\begin{equation}
  -\int_0^\tau\!\!\int_\Omega\qa\,\nabla\cdot(q\,\delta\bu)\dd\Omega\dd t
  =-\int_0^\tau\!\!\oint_{\partial\Omega}\qa\,q\,(\delta\bu\cdot\bn)\dd S\dd t
  +\int_0^\tau\!\!\int_\Omega q\,(\nabla\qa\cdot\delta\bu)\dd\Omega\dd t ,
\end{equation}
and one in $\delta\phi$,
\begin{align}
  \int_0^\tau\!\!\int_\Omega\qa\,\nabla\cdot(q\,\nabla\delta\phi)\dd\Omega\dd t
  = &\int_0^\tau\!\!\oint_{\partial\Omega}\qa\,q\,(\nabla\delta\phi\cdot\bn)\dd S\dd t\nonumber\\
  &-\int_0^\tau\!\!\oint_{\partial\Omega}q\,(\nabla\qa\cdot\bn)\,\delta\phi\dd S\dd t\nonumber\\
  &+\int_0^\tau\!\!\int_\Omega\nabla\cdot(q\,\nabla\qa)\,\delta\phi\dd\Omega\dd t ,
\end{align}
where the divergence theorem has been used twice.

\paragraph{Poisson's equation.}
Using Green's second identity,
\begin{align}
  -\int_0^\tau\!\!\int_\Omega\phia\left(\nabla^2\delta\phi+C\,\delta q\right)\dd\Omega\dd t
  =&-\int_0^\tau\!\!\int_\Omega\nabla^2\phia\,\delta\phi\dd\Omega\dd t\nonumber\\
  &+\int_0^\tau\!\!\oint_{\partial\Omega}\left(\nabla\phia\cdot\bn\,\delta\phi
    -\phia\,\nabla\delta\phi\cdot\bn\right)\dd S\dd t\nonumber\\
  &-C\int_0^\tau\!\!\int_\Omega\phia\,\delta q\dd\Omega\dd t .
\end{align}

\subsection{Complete first variation grouped by direct variable}

Collecting all the contributions above, and writing
$\bm{w}=\bu-\nabla\phi$ for the charge transport velocity, the first
variation of the Lagrangian is the sum of the contributions of the
variations of the four direct fields,
\begin{equation}
  \delta\mathcal{L}=\delta\mathcal{L}_{\bu}+\delta\mathcal{L}_{p}
  +\delta\mathcal{L}_{q}+\delta\mathcal{L}_{\phi},
  \label{eq:dL_grouped}
\end{equation}
each of which contains its terminal and initial, volume and boundary terms.

\paragraph{Velocity variation.} The terms in $\delta\bu$ are
\begin{align}
  \delta\mathcal{L}_{\bu}= &
  \int_\Omega\left[\bu(\tau)-\ua(\tau)\right]\cdot\delta\bu(\tau)\dd\Omega
  +\int_\Omega\ua(0)\cdot\delta\bu(0)\dd\Omega\nonumber\\
  &+\int_0^\tau\!\!\int_\Omega\delta\bu\cdot\Big[\partial_t\ua+(\bu\cdot\nabla)\ua
    -(\nabla\bu)^{\mathsf T}\cdot\ua+\nabla\pa+\nu\nabla^2\ua+q\nabla\qa\Big]
    \dd\Omega\dd t\nonumber\\
  &+\int_0^\tau\!\!\oint_{\partial\Omega}\Big\{\delta\bu\cdot\Big[-(\bu\cdot\bn)\ua
    -\nu\,\partial_n\ua-(\pa+q\qa)\,\bn\Big]
    +\nu\,\ua\cdot\partial_n\delta\bu\Big\}\dd S\dd t.
  \label{eq:dL_u}
\end{align}

\paragraph{Pressure variation.} The pressure enters only through the term
$\nabla\delta p$ of the momentum equation, which gives
\begin{equation}
  \delta\mathcal{L}_{p}=
  \int_0^\tau\!\!\int_\Omega\delta p\,\nabla\cdot\ua\dd\Omega\dd t
  -\int_0^\tau\!\!\oint_{\partial\Omega}(\ua\cdot\bn)\,\delta p\dd S\dd t .
  \label{eq:dL_p}
\end{equation}

\paragraph{Charge variation.}
With $\delta q(\bm{x},0)=\delta q'_0(\bm{x})$,
\begin{align}
  \delta\mathcal{L}_{q}=&
  -\int_\Omega\qa(\tau)\,\delta q(\tau)\dd\Omega
  +\int_\Omega\left[\qa(0)-2\lambda q'_0\right]\delta q'_0\dd\Omega\nonumber\\
  &+\int_0^\tau\!\!\int_\Omega\delta q\,\Big[\partial_t\qa+\bm{w}\cdot\nabla\qa
    -\beta\,\ua\cdot\nabla\phi-C\phia\Big]\dd\Omega\dd t
  -\int_0^\tau\!\!\oint_{\partial\Omega}\qa\,(\bm{w}\cdot\bn)\,\delta q\dd S\dd t .
  \label{eq:dL_q}
\end{align}

\paragraph{Potential variation.} The terms in $\delta\phi$ are
\begin{align}
  \delta\mathcal{L}_{\phi}=&
  \int_0^\tau\!\!\int_\Omega\delta\phi\,\Big[\beta\,\nabla\cdot(q\ua)
    +\nabla\cdot(q\nabla\qa)-\nabla^2\phia\Big]\dd\Omega\dd t\nonumber\\
  &+\int_0^\tau\!\!\oint_{\partial\Omega}\Big[(q\qa-\phia)\,\partial_n\delta\phi\nonumber\\
  &\qquad+\left(\partial_n\phia-q\,\partial_n\qa-\beta q\,\ua\cdot\bn\right)\delta\phi\Big]
    \dd S\dd t .
  \label{eq:dL_phi}
\end{align}

Each bracket multiplying a variation in the interior of $\Omega\times(0,\tau)$
gives one adjoint equation. Each boundary bracket gives the adjoint boundary
conditions compatible with the admissible variations of the direct fields.
The terminal terms give the terminal conditions of the backward
integration, and the remaining initial term gives the gradient with respect
to the control. These steps are carried out below. The same terms,
reorganized into volume and boundary contributions, are collected in
Eqs.~\eqref{eq:dL_volume}--\eqref{eq:dL_boundary}.

\subsection{Adjoint equations}

\paragraph{Volume and boundary contributions.}
The variation decomposes into volume and boundary contributions,
$\delta\mathcal{L}=\delta\mathcal{L}_\Omega+\delta\mathcal{L}_{\partial\Omega}$.
Writing $\bm{w}=\bu-\nabla\phi$ for the charge transport velocity, the volume
contribution grouped by variations of the direct fields is
\begin{align}
  \delta\mathcal{L}_\Omega= &
  \int_\Omega\left[\bu(\tau)-\ua(\tau)\right]\cdot\delta\bu(\tau)\dd\Omega
  +\int_\Omega\ua(0)\cdot\delta\bu(0)\dd\Omega\nonumber\\
  &+\int_0^\tau\!\!\int_\Omega\delta\bu\cdot\left[\partial_t\ua+(\bu\cdot\nabla)\ua
    -(\nabla\bu)^{\mathsf T}\cdot\ua+\nabla\pa+\nu\nabla^2\ua+q\nabla\qa\right]
    \dd\Omega\dd t\nonumber\\
  &+\int_0^\tau\!\!\int_\Omega\delta p\,\nabla\cdot\ua\dd\Omega\dd t\nonumber\\
  &-\int_\Omega\qa(\tau)\,\delta q(\tau)\dd\Omega
  +\int_\Omega\qa(0)\,\delta q(0)\dd\Omega
  -2\lambda\int_\Omega q'_0\,\delta q'_0\dd\Omega\nonumber\\
  &+\int_0^\tau\!\!\int_\Omega\delta q\left[\partial_t\qa+\bm{w}\cdot\nabla\qa
    -\beta\,\ua\cdot\nabla\phi-C\phia\right]\dd\Omega\dd t\nonumber\\
  &+\int_0^\tau\!\!\int_\Omega\delta\phi\left[\beta\,\nabla\cdot(q\ua)
    +\nabla\cdot(q\nabla\qa)-\nabla^2\phia\right]\dd\Omega\dd t .
  \label{eq:dL_volume}
\end{align}
The boundary contribution is
\begin{align}
  \delta\mathcal{L}_{\partial\Omega}=&
  \int_0^\tau\!\!\oint_{\partial\Omega}\Big[-(\bu\cdot\bn)(\ua\cdot\delta\bu)
    +\nu\,\ua\cdot\partial_n\delta\bu\nonumber\\
  &\qquad-\nu\,\delta\bu\cdot\partial_n\ua
    -\pa\,\delta\bu\cdot\bn-q\qa\,\delta\bu\cdot\bn\Big]\dd S\dd t\nonumber\\
  &-\int_0^\tau\!\!\oint_{\partial\Omega}(\ua\cdot\bn)\,\delta p\dd S\dd t
  -\int_0^\tau\!\!\oint_{\partial\Omega}\qa\,(\bm{w}\cdot\bn)\,\delta q\dd S\dd t\nonumber\\
  &+\int_0^\tau\!\!\oint_{\partial\Omega}\Big[(q\qa-\phia)\,\partial_n\delta\phi\nonumber\\
  &\qquad+\left(\partial_n\phia-q\,\partial_n\qa-\beta q\,\ua\cdot\bn\right)\delta\phi\Big]
    \dd S\dd t .
  \label{eq:dL_boundary}
\end{align}

\paragraph{Adjoint bulk equations.}
Since the variations of the direct fields are arbitrary in the interior of
$\Omega$, the coefficients of $\delta\bu$, $\delta p$, $\delta q$ and
$\delta\phi$ in Eq.~\eqref{eq:dL_volume} must vanish independently. This gives
the continuous adjoint equations
\begin{align}
  \frac{\partial\ua}{\partial t}+(\bu\cdot\nabla)\ua-(\nabla\bu)^{\mathsf T}\cdot\ua
    +\nabla\pa+\nu\nabla^2\ua+q\nabla\qa &= 0, \label{eq:adj_mom}\\
  \nabla\cdot\ua &= 0, \label{eq:adj_div}\\
  \frac{\partial\qa}{\partial t}+(\bu-\nabla\phi)\cdot\nabla\qa
    -\beta\,\ua\cdot\nabla\phi-C\phia &= 0, \label{eq:adj_charge}\\
  \beta\,\nabla\cdot(q\ua)+\nabla\cdot(q\nabla\qa)-\nabla^2\phia &= 0.
    \label{eq:adj_poisson}
\end{align}
Using $\bE=-\nabla\phi$, the last two equations can equivalently be written
as
\begin{align}
  \frac{\partial\qa}{\partial t}+(\bu+\bE)\cdot\nabla\qa+\beta\,\ua\cdot\bE-C\phia &= 0,\\
  \nabla^2\phia-\nabla\cdot\left[q\,(\nabla\qa+\beta\ua)\right] &= 0 .
\end{align}
Equations~\eqref{eq:adj_mom}--\eqref{eq:adj_poisson} are solved backward in
time, from $t=\tau$ to $t=0$, along the nonlinear direct trajectory.

\paragraph{Adjoint boundary conditions.}
The adjoint boundary conditions follow by requiring the boundary
contribution, Eq.~\eqref{eq:dL_boundary}, to vanish for every admissible variation
of the direct fields. At the electrodes the direct velocity satisfies no
slip, so $\delta\bu=0$ on $\Gamma_i\cup\Gamma_c$. Although $\delta\bu$
vanishes there, its normal derivative is arbitrary, and the viscous boundary
term requires
\begin{equation}
  \ua=\bm{0}\quad\text{on }\Gamma_i\cup\Gamma_c .
\end{equation}
In the periodic layer used in the computations, the contributions of the
two lateral boundaries cancel pairwise and no lateral adjoint conditions
arise. For a domain bounded by symmetry walls instead,
the direct symmetry conditions $u_x=0$,
$\partial u_y/\partial x=0$ imply $\delta u_x=0$,
$\partial\delta u_y/\partial x=0$, and substitution into the velocity
boundary terms gives the adjoint symmetry conditions
\begin{equation}
  u^{\dagger}_x=0,\qquad \frac{\partial u^{\dagger}_y}{\partial x}=0\quad\text{on }\Gamma_s .
\end{equation}
These conditions also make the boundary term in $\delta p$ vanish, since
$\ua\cdot\bn=0$ on $\partial\Omega$.

For the charge density, the direct value is prescribed at the injector, so
$\delta q=0$ on $\Gamma_i$. At the collector $\delta q$ is arbitrary. Since
the charge transport velocity $\bm{w}=\bu-\nabla\phi$ has a nonzero normal
component there, the charge boundary term requires
\begin{equation}
  \qa=0\quad\text{on }\Gamma_c .
\end{equation}
On the lateral symmetry boundaries $\bu\cdot\bn=0$ and $\partial_n\phi=0$, so
$\bm{w}\cdot\bn=0$. The charge boundary term then vanishes identically and no
condition on $\qa$ arises there.

The electric potential is prescribed at both electrodes, so
$\delta\phi=0$ on $\Gamma_i\cup\Gamma_c$, whereas $\partial_n\delta\phi$
remains arbitrary. The coefficient of $\partial_n\delta\phi$ in
Eq.~\eqref{eq:dL_boundary} must therefore vanish:
\begin{equation}
  \phia=q\qa\quad\text{on }\Gamma_i\cup\Gamma_c .
\end{equation}
In particular, since $\qa=0$ at the collector, $\phia=0$ on $\Gamma_c$. On the
lateral boundaries the direct condition is $\partial_n\phi=0$, hence
$\partial_n\delta\phi=0$, while $\delta\phi$ is arbitrary. The remaining
coefficient of $\delta\phi$ gives
$\partial_n\phia-q\,\partial_n\qa-\beta q\,\ua\cdot\bn=0$, which, with
$\ua\cdot\bn=0$, reduces to
\begin{equation}
  \partial_n\phia=q\,\partial_n\qa\quad\text{on }\Gamma_s .
\end{equation}

\paragraph{Terminal conditions.}
The terminal contributions in Eq.~\eqref{eq:dL_volume} are
\begin{equation}
  \int_\Omega\left[\bu(\tau)-\ua(\tau)\right]\cdot\delta\bu(\tau)\dd\Omega
  -\int_\Omega\qa(\tau)\,\delta q(\tau)\dd\Omega .
\end{equation}
Since $\delta\bu(\tau)$ and $\delta q(\tau)$ are arbitrary, the terminal
conditions for the time-dependent adjoint variables are
\begin{equation}
  \ua(\tau)=\bu(\tau),\qquad \qa(\tau)=0 .
\end{equation}
No terminal condition is required for $\phia$, since Eq.~\eqref{eq:adj_poisson} is
elliptic and is solved at each instant of the backward integration.
Similarly, $\pa$ acts as the Lagrange multiplier enforcing the adjoint
incompressibility constraint.

\paragraph{Gradient with respect to the initial charge perturbation.}
The initial velocity is fixed, and only the initial charge perturbation
$q'_0$ is used as control. Since $q(\bm{x},0)=q_b(\bm{x})+q'_0(\bm{x})$, we
have $\delta q(\bm{x},0)=\delta q'_0(\bm{x})$. After imposing the adjoint
bulk equations, boundary conditions and terminal conditions, the only
remaining first-order contribution associated with the control is
\begin{equation}
  \delta\mathcal{L}=\int_\Omega\left[\qa(\bm{x},0)-2\lambda q'_0(\bm{x})\right]
  \delta q'_0(\bm{x})\dd\Omega .
\end{equation}
The functional derivative of the augmented Lagrangian with respect to the
initial perturbation is therefore
\begin{equation}
  \frac{\delta\mathcal{L}}{\delta q'_0}=\qa(\bm{x},0)-2\lambda q'_0(\bm{x}).
\end{equation}
At a constrained stationary point this derivative vanishes, so that
$\qa(\bm{x},0)=2\lambda q'_0(\bm{x})$. The optimal initial charge
perturbation is parallel to the initial adjoint charge field. If the
amplitude constraint is treated separately by projection, the unconstrained
gradient of the reduced objective is simply
\begin{equation}
  g=\frac{\delta J}{\delta q'_0}=\qa(\bm{x},0),
\end{equation}
and only its component tangent to the constant-amplitude hypersphere
of Eq.~\eqref{eq:E0} is an admissible ascent direction,
\begin{equation}
  g_T=\qa(\bm{x},0)-\frac{\langle\qa(\bm{x},0),q'_0\rangle}{E_0}\,q'_0 .
\end{equation}

\section{Monolithic weak formulation of the direct problem}
\label{app:weak}

This appendix gives the discrete direct problem exactly as it is solved and
differentiated by the automatic adjoint (Sec.~\ref{sec:discretization}).

\paragraph{Function spaces.}
Let $\mathcal{T}_h$ be the structured quadrilateral mesh of the periodic
layer $\Omega=[0,L)\times[0,1]$, and $Q_k$ the space of continuous
piecewise polynomials of degree $k$ in each coordinate, periodic in $x$. The
perturbation fields are sought in
\begin{equation}
  \begin{aligned}
    \bu' &\in \bm{V}_h=\{\bv\in Q_2^2:\ \bv=\bm 0 \text{ on } \Gamma_i\cup\Gamma_c\}, &
    p' &\in Q_h=Q_1/\mathbb{R},\\
    q' &\in R_h=\{w\in Q_1:\ w=0 \text{ on } \Gamma_i\}, &
    \phi' &\in P_h=\{\psi\in Q_2:\ \psi=0 \text{ on } \Gamma_i\cup\Gamma_c\},
  \end{aligned}
  \label{eq:spaces}
\end{equation}
and $\bz'=(\bu',p',q',\phi')\in W_h=\bm{V}_h\times Q_h\times R_h\times P_h$.
The pressure is defined up to a constant, and this null space is removed from the
Newton and adjoint linear systems. The base state $\bz_b=(\bm 0,p_b,q_b,\phi_b)$ of
Eq.~\eqref{eq:base_nd}, with $p_b=CM^2E_i\,y$, is interpolated in the same spaces
(without boundary constraints), and the total fields are
\begin{equation}
  \bu=\bu',\qquad p=p_b+p',\qquad q=q_b+q',\qquad \phi=\phi_b+\phi' .
\end{equation}
Inner products are $(a,b)=\int_\Omega a\,b\dd\Omega$, and
$\nu=M^2/T$, $\beta=CM^2$ as in Appendix~\ref{app:adjoint}.

\paragraph{Time discretization.}
For a perturbation field $f$, the time derivative at level $n+1$ is
\begin{equation}
  \begin{gathered}
  D_tf^{n+1}=\frac{a_0f^{n+1}+a_1f^{n}+a_2f^{n-1}}{\Delta t},\\
  (a_0,a_1,a_2)=\begin{cases}(1,-1,0), & n=0 \text{ (backward Euler)},\\
  (\tfrac32,-2,\tfrac12), & n\ge1 \text{ (BDF2)}.\end{cases}
  \end{gathered}
  \label{eq:bdf}
\end{equation}
Since the base state is steady, $D_t$ acts on the perturbation only. Only
$\bu'$ and $q'$ carry a time derivative, and $p'$ and $\phi'$ are determined at
each level by the constraints.

\paragraph{Steady residual.}
For a total state $\bz=(\bu,p,q,\phi)$ and test functions
$(\bv,s,w,\psi)\in W_h$, define the charge transport velocity $\bv_q=\bu-\nabla\phi$,
the strong charge residual without time derivative
\begin{equation}
  r_q(\bz)=\bv_q\cdot\nabla q+C\,q^2 ,
  \label{eq:rq}
\end{equation}
which is the non-conservative form of Eq.~\eqref{eq:charge} after using
$\nabla\cdot\bE=Cq$ from Eq.~\eqref{eq:poisson}, and the SUPG parameter on each
cell $K$
\begin{equation}
  \tau_K(\bz)=\Big[\big(2|\bv_q|/h_K\big)^2+(2Cq)^2\Big]^{-1/2},
\end{equation}
with $h_K$ the largest distance between two vertices of $K$. The steady
residual is
\begin{multline}
  \mathcal{S}(\bz;\bv,s,w,\psi)=
  \big((\bu\cdot\nabla)\bu,\bv\big)+\nu(\nabla\bu,\nabla\bv)-(p,\nabla\cdot\bv)
  +\beta\,(q\nabla\phi,\bv)
  +(\nabla\cdot\bu,s)\\
  +(\nabla\phi,\nabla\psi)-C(q,\psi)
  +\big(r_q(\bz),w\big)
  +\sum_{K\in\mathcal{T}_h}\big(\tau_K\,r_q(\bz),\,\bv_q\cdot\nabla w\big)_K .
  \label{eq:steady_residual}
\end{multline}
The momentum terms are the weak form of Eq.~\eqref{eq:ns} with
$-\beta\,q\bE=\beta\,q\nabla\phi$. The boundary integrals vanish because
$\bv=\bm 0$ and $\psi=0$ on the electrodes and the lateral boundaries are
periodic. No boundary term appears in the charge equation, which is tested
in its strong form. The charge condition is imposed only at the injector,
and nothing is imposed at the collector, where the charge leaves the layer.

\paragraph{Time-step residual.}
At each level the perturbation $\bz'^{\,n+1}\in W_h$ satisfies, for all test
functions,
\begin{multline}
  \mathcal{F}(\bz'^{\,n+1};\bv,s,w,\psi)\equiv
  \big(D_t\bu'^{\,n+1},\bv\big)+\big(D_tq'^{\,n+1},w\big)
  +\sum_{K}\big(\tau_K\,D_tq'^{\,n+1},\,\bv_q\cdot\nabla w\big)_K\\
  +\mathcal{S}(\bz_b+\bz'^{\,n+1};\bv,s,w,\psi)
  -\mathcal{S}(\bz_b;\bv,s,w,\psi)=0 ,
  \label{eq:time_step_residual}
\end{multline}
where $\tau_K$ and $\bv_q$ in the time-derivative SUPG term are evaluated at
the total state $\bz_b+\bz'^{\,n+1}$, so that the SUPG term weights the full
strong residual $D_tq'+r_q$. Subtracting the base residual
$\mathcal{S}(\bz_b)$ makes $\bz'=0$ an exact discrete solution for any mesh
and time step (Sec.~\ref{sec:discretization}).

\paragraph{Nonlinear solution.}
Equation~\eqref{eq:time_step_residual} is solved for all unknowns at once by
Newton's method,
\begin{equation}
  \mathsf{J}\big(\bz'^{\,(k)}\big)\,\delta\bz=-\mathcal{F}\big(\bz'^{\,(k)}\big),
  \qquad \bz'^{\,(k+1)}=\bz'^{\,(k)}+\delta\bz ,
\end{equation}
where the Jacobian $\mathsf{J}$ is the exact G\^ateaux derivative of
Eq.~\eqref{eq:time_step_residual}, including the derivatives of $\tau_K$ and
$\bv_q$, assembled from the symbolic form. The iteration starts from the
previous time level and stops when the residual norm or the relative Newton
step falls below prescribed tolerances. The linear systems are solved by a
sparse direct factorization with null-pivot detection, and the constant
pressure mode is projected out.

\paragraph{Initial condition, objective and control.}
The integration starts from $\bu'^{\,0}=\bm 0$ and $q'^{\,0}=q'_0\in R_h$,
with $q'_0$ on the sphere $(q'_0,q'_0)=E_0$. After $N=\tau/\Delta t$ steps the
objective is
\begin{equation}
  J=\tfrac12\big(\bu'^{\,N},\bu'^{\,N}\big).
\end{equation}
The discrete adjoint is obtained by transposing the Jacobian of
Eq.~\eqref{eq:time_step_residual} at every level and solving backward in time. It
is therefore the exact derivative of $J$ with respect to $q'_0$ for this
discretization, including the SUPG and time-integration terms, rather than a
discretization of the continuous adjoint of Appendix~\ref{app:adjoint}. The
linearization of Eq.~\eqref{eq:steady_residual} at $\bz'=0$ is also the operator
of the marginal eigenvalue problem, Eq.~\eqref{eq:gevp}, after the rescaling
$p'\mapsto(T/M^2)\,p'$ that makes it linear in $T$.

\section{Transit-time modes}
\label{app:transit}

In this appendix we derive the dispersion relation, Eq.~\eqref{eq:transit_main}, for
horizontally uniform perturbations of the base state, the transit-time modes.
We show that these oscillatory modes are present even in the absence of fluid
flow.

Consider a small perturbation of the base state independent of $x$,
\begin{equation}
  \bu'=\bu'(y,t),\qquad p'=p'(y,t),\qquad q'=q'(y,t),\qquad \phi'=\phi'(y,t),
  \label{eq:tr_pert}
\end{equation}
and linearize the equations in the perturbation amplitude. The velocity perturbation is incompressible, $\nabla\cdot\bu'=0$, so continuity
gives $\partial_yu_y'=0$ and the wall conditions give $u_y'\equiv0$. The
horizontal velocity obeys
$\partial_tu_x'=(M^2/T)\partial_{yy}u_x'$, since neither the Coulomb force nor
the pressure gradient has an $x$-component. Its modes are the viscous shear modes
$\sigma=-(M^2/T)(m\pi)^2$, which the finite-element operator of
Sec.~\ref{sec:modal} reproduces to seven digits. The pressure balances the
perturbation of the Coulomb force. The charge is therefore not affected by the
flow, and the remaining problem is purely electrical,
\begin{equation}
  \partial_tq'+\partial_y\big(E_bq'+q_b E_y'\big)=0,\qquad \partial_y E_y'=Cq',\qquad
  q'(0,t)=0,\qquad \int_0^1E_y'\dd y=0 ,
  \label{eq:tr_1d}
\end{equation}
the last condition expressing that the applied voltage is fixed. For
$x$-independent perturbations the term $\partial_y(q'E_y')$ is the only
nonlinear term dropped, since $u_y'\equiv0$ holds exactly.

Eliminating $q'$ gives $\partial_y[\partial_tE_y'+C(E_bq'+q_bE_y')]=0$. The
perturbation of the total current, displacement plus conduction, is therefore
uniform across the layer,
\begin{equation}
  \partial_tE_y'+E_b\,\partial_yE_y'+Cq_b\,E_y'=I(t),
  \label{eq:tr_current}
\end{equation}
where $I(t)$, proportional to the perturbation of the total current, takes the
value that keeps the voltage fixed. At the injector $q'=0$ gives $\partial_yE_y'=0$, so
\begin{equation}
 \dot E_i'+CE_i'=I, 
 \label{eq:tr_injector}
\end{equation}
where $E_i'(t)=E_y'(0,t)$ is the perturbation of the electric field at the
injector. The drift time for an ion to go from the injector to a point at height
$y$ is
\begin{equation}
  \tau(y)=\int_0^y\frac{\dd s}{E_b(s)}=
  \dfrac{1}{C}\left(\sqrt{1+2C y/E_i} -1\right),
  \label{eq:tr_tau}
\end{equation}
and the transit time is $t_{\rm tr}=\tau(1)=C^{-1}(\sqrt{1+2C/E_i}-1)$.
The base state expressed in terms of $\tau$ is
\begin{equation}
  y=E_i\left(\tau+\frac{C\tau^2}{2}\right), \qquad
  E_b=E_i(1+C\tau),\qquad 
  q_b=\frac1{1+C\tau}.
  \label{eq:tr_base}
\end{equation}
Since $\partial_\tau=E_b\partial_y$, Eq.~\eqref{eq:tr_current} becomes
$(\partial_t+\partial_\tau)E_y'+CE_y'/(1+C\tau)=I$. The operator
$\partial_t+\partial_\tau$ is the derivative following an ion that moves with
velocity equal to $E_b$. For this ion it is $\dd\tau/\dd t=1$. With the
integrating factor $1+C\tau$, the function $F=(1+C\tau)E_y'$ satisfies
\begin{equation}
  (\partial_t+\partial_\tau)F=(1+C\tau)\,I(t),\qquad F(0,t)=E_i'(t),\qquad
  \int_0^{t_{\rm tr}}F\dd\tau=0 .
  \label{eq:tr_F}
\end{equation}
The integral condition comes from the last condition in Eq.~\eqref{eq:tr_1d}, with
the change $\mathrm{d}y=E_b\mathrm{d}\tau$ and using the expression for $E_b$ in
Eq.~\eqref{eq:tr_base}. 

Equations~\eqref{eq:tr_F} and \eqref{eq:tr_injector} form a closed system for
the unknowns $\{F(\tau,t), E_i'(t), I(t)\}$. We seek normal modes of the form
\begin{equation}
(F,E_i',I)=(\hat F(\tau),\hat{E}_i',\hat I)\,e^{\sigma t}.
\label{eq:tr_normal}
\end{equation}
The system becomes
\begin{equation}
\begin{aligned}
\hat F'(\tau)+\sigma\hat F(\tau)&=(1+C\tau)\,\hat I, \\
\hat I&=(\sigma+C)\hat{E}_i',\\
\int_0^{t_{\rm tr}}\hat F\dd\tau&=0.
\end{aligned}
\label{eq:tr_modes}
\end{equation}
Multiplying the first equation by $e^{\sigma\tau}$ and integrating from $0$ to $\tau$ gives
\begin{equation}
\hat F(\tau)=\hat{E}_i'\,e^{-\sigma\tau}+\hat I\int_0^\tau(1+Cs)e^{-\sigma(\tau-s)}\dd s.
\label{eq:tr_F_int}
\end{equation}
We introduce this expression into the integral condition to obtain
\begin{equation}
A(\sigma)\hat{E}_i'+B(\sigma)\hat I=0,
\label{eq:tr_integral}
\end{equation}
with
\begin{align}
A(\sigma)&=\int_0^{t_{\rm tr}}e^{-\sigma\tau}\dd\tau = \frac{1-e^{-t_{\rm tr}\sigma}}{\sigma},
\label{eq:tr_A}\\
B(\sigma)&=\int_0^{t_{\rm tr}}\int_0^\tau(1+Cs)e^{-\sigma(\tau-s)}\dd s\dd\tau\nonumber\\
&=\frac{-1+e^{-t_{\rm tr}\sigma}+t_{\rm tr}\sigma}{\sigma^2}
+C\,\frac{2-2e^{-t_{\rm tr}\sigma}+t_{\rm tr}\sigma(-2+t_{\rm tr}\sigma)}{2\sigma^3}.
\label{eq:tr_B}
\end{align}
Now the system becomes
\begin{equation}
  \left(
  \begin{array}{cc}
    A(\sigma) & B(\sigma) \\
    \sigma+C & -1
  \end{array}
  \right)
  \left(
    \begin{array}{c}
      \hat{E}_i' \\
      \hat I
    \end{array}
  \right)
  =
  \left(
    \begin{array}{c}
      0 \\
      0
    \end{array}
  \right).
\label{eq:tr_system}
\end{equation}
For this system to have a nontrivial solution, the determinant must vanish, which gives
\begin{equation}
A(\sigma)+B(\sigma)(\sigma+C)=0.
\label{eq:tr_D}
\end{equation}
From Eq.~\eqref{eq:tr_base}, evaluating at $y=1$, we have
\begin{equation}
  t_{\rm tr} + \dfrac{1}{2}Ct_{\rm tr}^2 = \dfrac{1}{E_i}, \qquad 1 + Ct_{\rm tr} = \dfrac{E_c}{E_i},
\end{equation}
where $E_c=E_b(1)$ is the collector field. Multiplying Eq.~\eqref{eq:tr_D} by
$\sigma^3$ and using these relations, we rewrite the dispersion
relation as
\begin{equation}
  P(\sigma) = C^2 e^{-\sigma t_{\rm tr}},
  \label{eq:tr_main}
\end{equation}
with 
\begin{equation}
  P(\sigma)=\sigma^2+(\sigma+C)\Big[\frac{\sigma^2}{E_i}-\frac{E_c}{E_i}\,\sigma+C\Big].
  \label{eq:tr_P}
\end{equation}
The factor $\sigma^3$ introduces a triple root $\sigma=0$ of
Eq.~\eqref{eq:tr_main}, which is spurious. Indeed $A(0)=t_{\rm tr}$ and
$B(0)=t_{\rm tr}^2/2+Ct_{\rm tr}^3/6$ are finite and positive, so $\sigma=0$ is
not a root of Eq.~\eqref{eq:tr_D}.
The cubic $P(\sigma)$ is the response of the layer if the disturbance were lost
at the collector. The right-hand side of Eq.~\eqref{eq:tr_main} is the part
that returns, through the voltage constraint, after one transit.

Because $e^{-\sigma t_{\rm tr}}$ is periodic in the imaginary direction with
period $2\pi\mathrm i/t_{\rm tr}$, Eq.~\eqref{eq:tr_main} is equivalent to
$-\sigma t_{\rm tr}=\mathrm{Ln}\big(P(\sigma)/C^2\big)-2\pi\mathrm im$ for some
integer $m$, where $\mathrm{Ln}$ is the principal branch of the logarithm.
Solving for $\sigma$ gives one equation for each branch,
\begin{equation}
  \sigma_m=-\frac{1}{t_{\rm tr}}\left[\mathrm{Ln}\left(\frac{P(\sigma_m)}{C^2}\right)-2\pi\mathrm im\right],
  \qquad m=0,\pm1,\pm2,\dots
  \label{eq:tr_log}
\end{equation}
The coefficients of $P$ are real, so the branch $-m$ gives the complex conjugate
of $\sigma_m$ and it suffices to consider $m\geq0$. Each branch with $m\geq1$ gives one
transit mode, and the branch $m=0$ gives only the spurious root $\sigma=0$.
Since the logarithm varies slowly, its derivative $P'/P$ being small compared
with $t_{\rm tr}$, consecutive roots are separated by
approximately $2\pi\mathrm i/t_{\rm tr}$, as expected for a disturbance that
returns once per transit.

For $C=10$ the first roots, computed by Newton iteration on
Eq.~\eqref{eq:tr_D} and checked with the argument principle, are compared with the
$n=0$ modes of the spectral code in Table~\ref{tab:transit}.

\begin{table*}
  \centering
  \caption{Horizontally uniform transit-time modes for $C=10$ ($t_{\rm
    tr}=1.248$, $2\pi/t_{\rm tr}=5.04$). Roots of Eq.~\eqref{eq:tr_D} and $n=0$
    eigenvalues of the spectral code of Sec.~\ref{sec:transit} for $N=60$ and of
    the finite-element operator of Sec.~\ref{sec:modal}, which are the same for
    every $M$ and $T$.}
  \label{tab:transit}
  \begin{tabular}{cccc}
    \toprule
    $m$ & roots of Eq.~\eqref{eq:tr_D} & spectral, $N=60$ & finite elements\\
    \midrule
    1 & $-3.01437\pm6.09207\,\mathrm i$ & $-3.01448\pm6.09201\,\mathrm i$ & $-3.034\pm6.100\,\mathrm i$ \\
    2 & $-4.09452\pm11.06757\,\mathrm i$ & $-4.09478\pm11.06757\,\mathrm i$ & $-4.144\pm11.126\,\mathrm i$\\
    3 & $-4.88280\pm16.07903\,\mathrm i$ & $-4.88317\pm16.07912\,\mathrm i$ & $-4.974\pm16.222\,\mathrm i$\\
    \bottomrule
  \end{tabular}
\end{table*}

\end{widetext}
\FloatBarrier

\bibliographystyle{aipnum4-2}
\bibliography{pofcontrol,paper}

\end{document}